\UseRawInputEncoding
\documentclass[aps,prl,twocolumn,showpacs,superscriptaddress,floats]{revtex4-1}
\usepackage{amssymb}
\usepackage{amsmath}
\usepackage{bm}
\usepackage{epsfig}
\usepackage{bbold}
\usepackage{graphicx}
\usepackage[colorlinks=true,linkcolor=blue]{hyperref}
\usepackage{color}
\definecolor{blue}{rgb}{0,0.0,0.9}
\definecolor{green}{rgb}{0.0,0.9,0.0}

\usepackage{epstopdf}
\newcommand{\Ztwo}{\mathbf{Z}_2}

\begin{document}

\title{Stable and unstable trajectories in a dipolar chain}

\author{Jaime Cisternas}
\email[]{jecisternas@miuandes.cl}
\affiliation{Complex Systems Group, Facultad de Ingenier\'{\i}a y Ciencias Aplicadas,
Universidad de los Andes, Chile, Alvaro del Portillo 12455, Santiago, Chile}
\author{Paula Mellado}
\affiliation{Condensed Matter i-Lab,  Adolfo Ib\'a\~{nez} University, Diagonal las torres 2640, Building D, Pe\~{n}alol\'en, Santiago, Chile}
\affiliation{School of Engineering and Sciences, Universidad Adolfo Ib\'a\~{n}ez, Diagonal las torres 2640, Pe\~{n}alol\'en, Santiago, Chile} 
\author{Felipe Urbina}
\affiliation{Condensed Matter i-Lab,  Adolfo Ib\'a\~{nez} University, Diagonal las torres 2640, Building D, Pe\~{n}alol\'en, Santiago, Chile}
\affiliation{Centro de Investigaci\'on DAiTA Lab, Facultad de Estudios Interdisciplinarios, Universidad Mayor, Santiago, Chile}
\author{Crist\'obal Portilla}
\affiliation{Condensed Matter i-Lab,  Adolfo Ib\'a\~{nez} University, Diagonal las torres 2640, Building D, Pe\~{n}alol\'en, Santiago, Chile}
\affiliation{School of Engineering and Sciences, Universidad Adolfo Ib\'a\~{n}ez, Diagonal las torres 2640, Pe\~{n}alol\'en, Santiago, Chile}
\author{Miguel Carrasco}
\affiliation{Condensed Matter i-Lab,  Adolfo Ib\'a\~{nez} University, Diagonal las torres 2640, Building D, Pe\~{n}alol\'en, Santiago, Chile}
\affiliation{School of Engineering and Sciences, Universidad Adolfo Ib\'a\~{n}ez, Diagonal las torres 2640, Pe\~{n}alol\'en, Santiago, Chile}
\author{\\Andr\'es Concha}
\email[]{andres.physics.research@gmail.com}
\affiliation{Condensed Matter i-Lab,  Adolfo Ib\'a\~{nez} University, Diagonal las torres 2640, Building D, Pe\~{n}alol\'en, Santiago, Chile}
\affiliation{School of Engineering and Sciences, Universidad Adolfo Ib\'a\~{n}ez, Diagonal las torres 2640, Pe\~{n}alol\'en, Santiago, Chile}
  
\begin{abstract}
In classical mechanics, solutions can be classified according to their stability. Each of them is part of the possible trajectories of the system. However, the signatures of unstable solutions are hard to observe in an experiment, and most of the times if the experimental realization is adiabatic, they are considered just a nuisance. Here we use a small number of XY magnetic dipoles subject to an external magnetic field for studying the origin of their collective magnetic response. Using bifurcation theory we have found all the possible solutions being stable or unstable, and explored how those solutions are naturally connected by points where the symmetries of the system are lost or restored. Unstable solutions that reveal the symmetries of the system are found to be the culprit  that shape hysteresis loops in this system. The complexity of the solutions for the nonlinear dynamics is analyzed using the concept of boundary basin entropy, finding that the damping time scale is critical for the emergence of fractal structures in the basins of attraction. Furthermore, we numerically found domain wall solutions that are the smallest possible realizations of transverse walls and vortex walls in magnetism. We experimentally confirmed their existence and stability showing that our system is a suitable platform to study domain wall dynamics at the macroscale.   
\end{abstract}

\maketitle

\section{Introduction}
One of the most important tasks in material science has been the search, discovery, and design of materials that can display memory effects, i.e., that are able to store information and allow for reading and writing cycles \cite{handley2000modern,coey2010magnetism,merz1953double,woodward1960particle,concha2018designing}. In this quest, the study of magnetic materials, models such as Ising \cite{ising1925beitrag,onsager1944crystal,kaufman1949crystal} and  spin glasses have been introduced. Spin glasses \cite{randeria1985low,sethna1993hysteresis} and other models have proved to be useful to understand the building  blocks of hysteresis loops in terms of hysterons, a phenomenological quanta of irreversibility \cite{jacobs1957magnetic,woodward1960particle,jiles1986theory,krasnosel2012systems}, and most recently the role of an exponentially large number of states have shown new phenomena in spin ices \cite{ramirez1999zero,castelnovo2008magnetic}.
Even though the one dimensional Ising model lacks long range order owing to the generation of domain wall excitations, it has become part of the useful models to understand as diverse phenomena as magnetism, phase transitions, or even social segregation \cite{domic2011dynamics}. Beyond the simplest Ising model, the properties of linear chains with non-Ising spin coupling are of broad significance. There are several materials which are known to behave as quasi-1D spin systems exhibiting either ferromagnetic or antiferromagnetic behavior. For higher spin values or in the semi-classical regime, ethylammonium manganese trichloride (TMMC) is a quasi-1D spin $5/2$, with an anisotropy into the planar XY regime \cite{kamminga2018out}. 

Experiments have now widely evidenced the relevance of dipolar interactions.  In the A2B2O7 pyrochlore oxides \cite{gardner2010magnetic}, dipolar interactions can be appreciable. This is also the case of nanomagnetic arrays, collections of nanomagnetic islands arranged in a regular pattern using lithography \cite{nisoli2013colloquium}. The magnitude of the moments as well as the strength of the dipolar interactions can be tuned by controlling the dimensions and separation of the magnetic islands. Polar molecules and atomic gases with large dipole moments confined in optical lattices and organic chains are new examples of quasi-1D dipolar systems \cite{pupillo2009condensed,peter2012anomalous}. 
When combining low dimensionality, dipolar interactions and XY anisotropy, the subject of our interest, theoretical results are scarce, and limited to the study of a few thermodynamic and dynamical aspects. Indeed Monte Carlo simulations of the 1-D XY model with power law decay interactions $r^{(1+\sigma)} $in the case of $\sigma= 1$, conveyed that the specific heat has no system size dependence and confirmed the theoretical prediction of the existence of some  `Berezinskii-Kosterlitz Thouless-like' transition with the exponent $\eta = 1$ \cite{kadanoff2000statistical}.

\begin{figure*}[t]
\begin{center}
\includegraphics[width=1.0\textwidth]{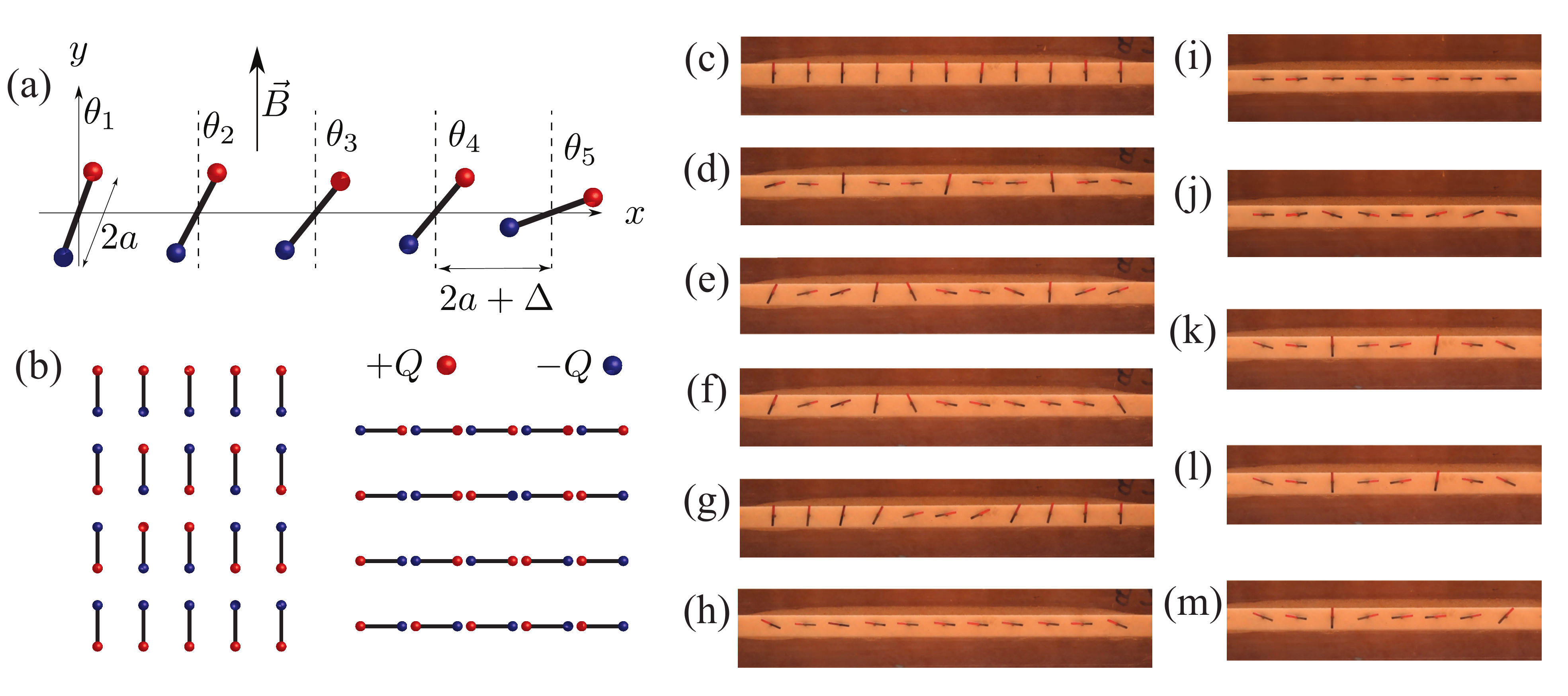}
\end{center}
\caption{(a) Schematic of linear array of $N=5$ magnetic dipoles.
Dipoles (depicted here with positive charges in red and negative charges in blue)
can rotate around hinges that are equally spaced and define a line perpendicular to the external magnetic field. 
(b) Configurations of parallel dipoles are trivial static solutions, independent of the external field,
but they are unstable with the exception of the $\theta_\alpha=0$ (all-up), $\theta_\alpha=\pi$ (all-down), $\theta_\alpha=\pi/2$ (all-right), and $\theta_\alpha=-\pi/2$ (all-left).
Here we show a limited selection of representative configurations. 
(c-m) Experimental realization of the model showing fully polarized configuration (c) and a ferromagnetic ground state (i). (d-f) and (j-m) stable defects found by programming the initial configuration of the system. Experimental chain sizes were $N=11$ and $N=8$.}
\label{fig:dumbbells}
\end{figure*}

It was experimentally  shown \cite{concha2018designing}, using a macroscopic dipolar chain (See Fig.\ref{fig:dumbbells}), that depending on the size and separation between dipoles, a variety of stable static magnetic configurations: 1) parallel to the external magnetic field; 2) aligned with each other; or 3) `canted' may show up. All of these are such that a balance between the internal dipolar interactions and the external field is reached \cite{mellado2012macroscopic,concha2018designing}. Furthermore, it was demonstrated that hysteresis loops can be designed in this system by controlling its geometry or the interaction strength \cite{concha2018designing}.  That particular experimental setup is extremely flexible as high quality magnets can be easily obtained in a variety of magnetic strength, length and mass combinations.
 
Those previously found states may coexist, and when the external field is slowly varied, can give rise to hysteresis regions with features such as: shoulders in the hysteresis loops, sub-critical, and super-critical transitions,  that have not been explained up to date. These findings could not be explained by a point-like model for the dipoles (as treated by classic electromagnetism textbooks such as \cite{pollack2002electromagnetism}). That is clearly an oversimplification in the case where dipoles are not far away with respect to each other. Instead, we use a finite model that assumes that the separation between dipoles is of the same order as their size (See Fig.\ref{fig:dumbbells}). This assumption allows the study of the effect of separation between them, and transitions between monopolar and dipolar regimes.

The main objective of this article is to study the birth, evolution, and death of all the possible solutions allowed by this classical system. We unveil the existence of unstable solutions and nontrivial stable solutions (See Fig.\ref{fig:dumbbells} (c-m)) and characterize the symmetries of all the states. We use bifurcation theory and numerical continuation for an exhaustive exploration of all the branches. We find spontaneous symmetry-breaking bifurcations, stable `localized' solutions, and when several stable states coexist, nontrivial basins of attraction with fractal boundaries. We also show that these stable `localized' solutions are the smallest possible realizations of domain walls (DW) reported long ago in micromagnetic simulations of permalloy films  \cite{mcmichael1997head}, and nicely explained in terms of emergent topological defects \cite{tchernyshyov2005fractional,clarke2008dynamics}. To test our numerical findings, we use an experimental setup similar to the one used in reference \cite{concha2018designing}  to  confirm the existence of several stable states that contain defects for certain values of the experimental  parameters. Our findings provide the simplest realization of DW, showing that dipolar objects at any scale are perfect candidates to support this type of topological defects. Our work opens up several venues for future research, such as the study of resonant motion of DW \cite{tretiakov2010minimization}, the stabilization of unstable trajectories \cite{kapitza1951dynamic}, the study of magneto-elastic metamaterials \cite{bar2020geometric,florijn2014programmable}, and the fast evaluation of path integrals for small magnetic clusters \cite{RevModPhys.20.367,kleinert2009path}.  
\section{Model and symmetries}
Since the length of the dipoles is relevant, we use a `dumbbell' model that
assumes that the charges are concentrated at the endpoints of bars of uniform density
(see Fig.\ref{fig:dumbbells} for a schematic and the definition of the angles).
For each dipole (labeled by $\alpha=1\ldots N$; charges belonging to a given dipole are labeled by $i \in \alpha$)
we consider the four torques induced by every other dipole
(full long-range Coulomb interaction between the two pairs of charges in the two dipoles):
\begin{multline}
I_\alpha \frac{d^2 \theta_\alpha}{dt^2} = \left( \frac{\mu_0}{4 \pi} \right)
\sum_{\beta \neq \alpha} \sum_{i \in \alpha} \sum_{j \in \beta}
Q_i Q_j \left( \vec{a}_i \times \frac{\vec{r}_{ij}}{r_{ij}^3}\right)\cdot \hat{z} \\
- \eta_\alpha \frac{d \theta_\alpha}{dt} + \left( \vec{P}_\alpha \times \vec{B} \right)\cdot \hat{z}
\label{model}
\end{multline}
where $I_\alpha$ is the moment of inertia of the dipole,
$\mu_0$ is the vacuum permeability,
$Q_i$ is the magnetic charge,
$\vec{a}_i$ is the vector that goes from the rotation center to the charge $Q_i$,
$\vec{r}_{ij}$ is the vector that goes from the position of $Q_j$ to the position of $Q_i$,
and $\eta_\alpha$ is the damping. The external magnetic field $\vec{B} = B \hat{y}$ is uniform.

The magnetic dipolar moment of each dipole is
\begin{equation}
\vec{P}_\alpha = \sum_{i \in \alpha} Q_i \vec{a}_i \hspace{0.5cm}.
\end{equation}

Unless explicitly stated, we assume identical dipoles:
$|Q_i|=Q, |\vec{a}_i|=a, I_\alpha=I, \eta_\alpha=\eta$,
and centers of rotation placed in a linear lattice of separation $2a+\Delta$.
This linear array is oriented perpendicular to the applied field $\vec{B}$. 

Given that there are several terms competing in Eq.(\ref{model}) it is useful to find the typical time scales at play. We will use them as control parameters in this work, and dimensionless  quantities represent ratios of those time scales. The magneto-inertial time scale is
\begin{equation}
\tau_{B}=2\pi \left( \frac{I}{2aQB} \right)^{1/2}  \hspace{0.5cm}.
\end{equation} 
The damping time scale is
\begin{equation}
\tau_{\eta}=I/\eta  \hspace{0.5cm}.
\end{equation}
The Coulombic time scale is
\begin{equation}
\tau_{c}=  \left( \frac{4\pi \Delta^2 I}{\mu_{0} a Q^2} \right)^{1/2}\hspace{0.5cm},
\end{equation} 
where the closest possible distance between magnetic charges $\Delta$ has been used as the relevant length scale for this problem. Full details of the definition of these time scales can be found in the Supplementary Information of reference \cite{concha2018designing}. 

Finally, collective oscillations of the ferromagnetic ground state ($\vec{B}=\vec{0}$) generate collective normal modes. The dispersion relation for normal modes is:
\begin{eqnarray}
\omega=\sqrt{\left(\frac{\mu_{0}}{4\pi}\right)\frac{a^2 Q^2}{I\Delta^3}\left[1+\cos\left(k L\right)\right]} \hspace{0.5cm},
\end{eqnarray}
that in the low energy limit $k L\rightarrow 0$ defines a collective time scale given by 
\begin{eqnarray}
\tau_{o}=2\pi\sqrt{\left(\frac{2\pi}{\mu_{0}}\right)\frac{I\Delta^3}{a^2 Q^2}}
\end{eqnarray}
which corresponds to the typical time scale for an excitation that has a canted magnetic texture. That is, all angular displacements in phase have $ \theta_\alpha \sim 1$ . This type of oscillations can be excited using a magnetic  
field of the type $\vec{B}(t)=B_{0}\sin(\Omega t) \hat{y}$. The lowest energy excitation mode is the one corresponding 
to $k L\rightarrow \pi$ corresponding to a spin-wave excitation, that is $\theta_\alpha \sim (-1)^{\alpha}$. 
The ratio between  the time scales $\tau_{o}$ and $\tau_{c}$ shows that a relevant dimensionless parameter for this system is $\Delta/ a$. Therefore, we can use as control parameters $\Delta$, that is the minimum distance between charges of neighboring dipoles,
and $b=B/B_c$ which is a dimensionless magnetic field, where $B_c=(\mu_0/4\pi)(Q/a^2)$
is a characteristic internal field.

For characterization of the states of the dipoles we use the averages of the horizontal and vertical projections of the long magnetization axes:
\begin{eqnarray}
m_x &=& \frac{1}{N} \sum_{\alpha=1}^N \sin \theta_\alpha  \hspace{0.5cm},\\
m_y &=& \frac{1}{N} \sum_{\alpha=1}^N \cos \theta_\alpha \hspace{0.5cm}.
\end{eqnarray}
In the following, we will use two complementary diagrams: $(b, m_y)$ and $(b, m_x)$.
Because of the symmetries of the system, a point in these diagrams may represent more than one solution configuration of Eq.~(\ref{model}).  

\subsection{System symmetries}

A linear chain of rotating dumbbells has certain symmetries that are
relevant for the existence of trivial solutions and for the degeneracy of
nontrivial ones \cite{golubitsky2012singularities}. Group Theory offers a rigorous characterization of the symmetries of
the solutions and the symmetry-breaking bifurcations that can be generically expected.

A symmetry of the system is a transformation of the angles $\theta = \{ \theta_1, \theta_2,\ldots \theta_N \}$
that appears as a regular transformation of the torques:
$$
\gamma F(\theta;b) = F(\gamma(\theta;b)) \hspace{0.5cm},
$$
where $F$ is the right hand side of Eq.~(\ref{model}) assuming $\dot{\theta}=0$.
These transformations map every equilibrium state $\theta$ onto
equilibrium states with the same stability.

More concretely, the dynamics of the linear array of dipoles is invariant under:
\begin{itemize}
\item a reflection with respect to vertical axis through the midpoint of the linear array:
$$
\nu : \theta_\alpha \rightarrow -\theta_{N-\alpha+1} \hspace{0.5cm}.
$$
\end{itemize}
A solution $\theta$ and its transformation $\nu \theta$ will share the same value of $m_y$,
but $m_x$ with opposite sign.
This transformation satisfies $\nu^2=1$ and generates the cyclic group $\Ztwo(\nu)$.

\begin{itemize}
\item a reflection of each dipole with respect to the vertical direction: 
$$
\kappa : \theta_\alpha \rightarrow -\theta_\alpha
$$
\end{itemize}
A solution $\theta$ and its transformation $\kappa \theta$
will share the same $m_y$, but $m_x$ with opposite sign.
This transformation generates the cyclic group $\Ztwo(\kappa)$:
\begin{itemize}
\item a field reversal: 
$$
\mu : b \rightarrow -b , \theta_\alpha \rightarrow \pi + \theta_\alpha \hspace{0.5cm}.
$$
\end{itemize}
The whole diagram $(b, m_y)$ will be symmetric under a reflection through the center,
and the diagram $(b,m_x)$ will be symmetric under reflections with respect to the vertical and horizontal axes.
Unless $b=0$, this transformation is a \emph{parameter symmetry} since it involves both the state $\theta$ and the parameter $b$;
it does not affect the multiplicity of solutions but induces a symmetry in the bifurcation diagrams.

\begin{figure}
\begin{center}
\includegraphics[width=0.3\textwidth]{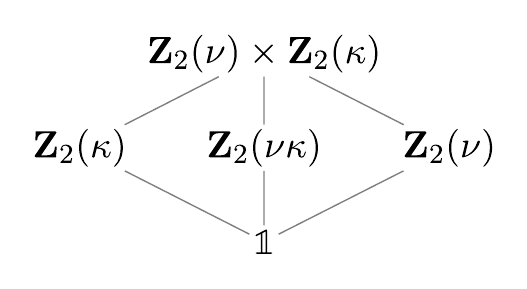}
\end{center}
\caption{Isotropy lattice of $\Ztwo(\nu) \times \Ztwo(\kappa)$.
}
\label{fig:lattice}
\end{figure}

The symmetry of the generic system $b \ne 0$ is characterized by the group $\Gamma = \Ztwo(\nu) \times \Ztwo(\kappa)=\{1,\nu,\kappa,\nu \kappa \}$.
We will classify the symmetry of a solution $\theta$ by its isotropy subgroup $\Sigma_\theta = \{ \gamma \in \Gamma | \gamma\theta=\theta \}$.
For the linear array, the possible isotropy subgroups are $\Ztwo(\nu) \times \Ztwo(\kappa)$, $\Ztwo(\nu \kappa)$, $\Ztwo(\nu)$, $\Ztwo(\kappa)$ and the trivial group $\mathbb{1}$,
as depicted in Fig.\ref{fig:lattice}. This diagram predicts possible symmetries of solutions (described by their isotropy subgroups).
Connections between subgroups represent possible symmetry-breaking bifurcations as parameters of the system are varied.
For instance with $N=3$ and small $\Delta$, the up-up-up solution, that has symmetry $\Ztwo(\nu) \times \Ztwo(\kappa)$, bifurcates for some value of $b$ to the left-up-right solution of symmetry $\Ztwo(\nu)$,
that in turn suffers a secondary bifurcation at some other value of $b$ where it loses the remaining symmetry.
Now some of the links in the isotropy lattice are not associated to bifurcations,
since the system does not allow continuous deformations with the required symmetries,
as is the case between configurations up-up-up (isotropy subgroup $\Ztwo(\nu)\times \Ztwo(\kappa)$)
and up-up-down (isotropy subgroup $\Ztwo(\kappa)$).
The isotropy lattice does not predict the stability of the solutions or the precise location of the bifurcations in parameter space.
In the supplementary information, we apply the trace formula to
a representation of these transformations and obtain predictions for
the dimensions of the fixed spaces.

\subsection{Trivial static states of the linear array}

The symmetries of the system allow the identification of simple configurations that are trivial static equilibria.

All the $2^N$ configurations with dipoles parallel to the (nonzero) magnetic
field, $\theta_\alpha = 0$ or $\pi$ (see Fig.~\ref{fig:dumbbells}(b)) are static solutions
for arbitrary field intensity $b$.
These are the parallel trivial solutions that satisfy $\kappa \theta = \theta$.
They are all unstable saddle configurations with the exception of the fully polarized states:
\begin{itemize}
\item $\theta_\alpha = 0$, ``all-up'', stable for large positive $b$;
\item $\theta_\alpha = \pi$, ``all-down'', stable for large negative $b$.
\end{itemize}
All parallel configurations are invariant under action $\kappa$,
but only a few are invariant under $\nu$.

Some of the parallel configurations can be connected via a transformation in $\Gamma$
(they belong to the same \emph{group orbit}, for instance up-up-down and down-up-up)
and some of the values of $m_y$ or $m_x$ (or their absolute values) may coincide.

Some of the parallel configurations do coincide in their values of $m_y$ and $m_x$
despite having different symmetries, for instance up-down-up and up-up-down. 
They are not connected via a transformation in $\Gamma$.

In the absence of an external magnetic field $b=0$, all the $2^N$
configurations with aligned dipoles $\theta_\alpha = \pm \pi/2$ (see Fig.~\ref{fig:dumbbells}(b))
are static solutions.
These are the aligned trivial solutions.
They are unstable with the exception of the ferromagnetic states:
\begin{itemize}
\item $\theta_\alpha = \pi/2$, ``all-right'', stable for $b=0$;
\item $\theta_\alpha = -\pi/2$, ``all-left'', stable for $b=0$.
\end{itemize}
These are the aligned trivial solutions.
Some aligned configurations are invariant under actions $\kappa$ or $\kappa\nu$.

Now for nonzero $b$ the solution branches that result from continuation of these solutions 
retain some of the symmetries of the initial solution and may switch stability, as
we will show in the next section. These bifurcations may be associated to merging of branches.


\begin{figure*}[t]
\begin{center}
\includegraphics[width=\textwidth]{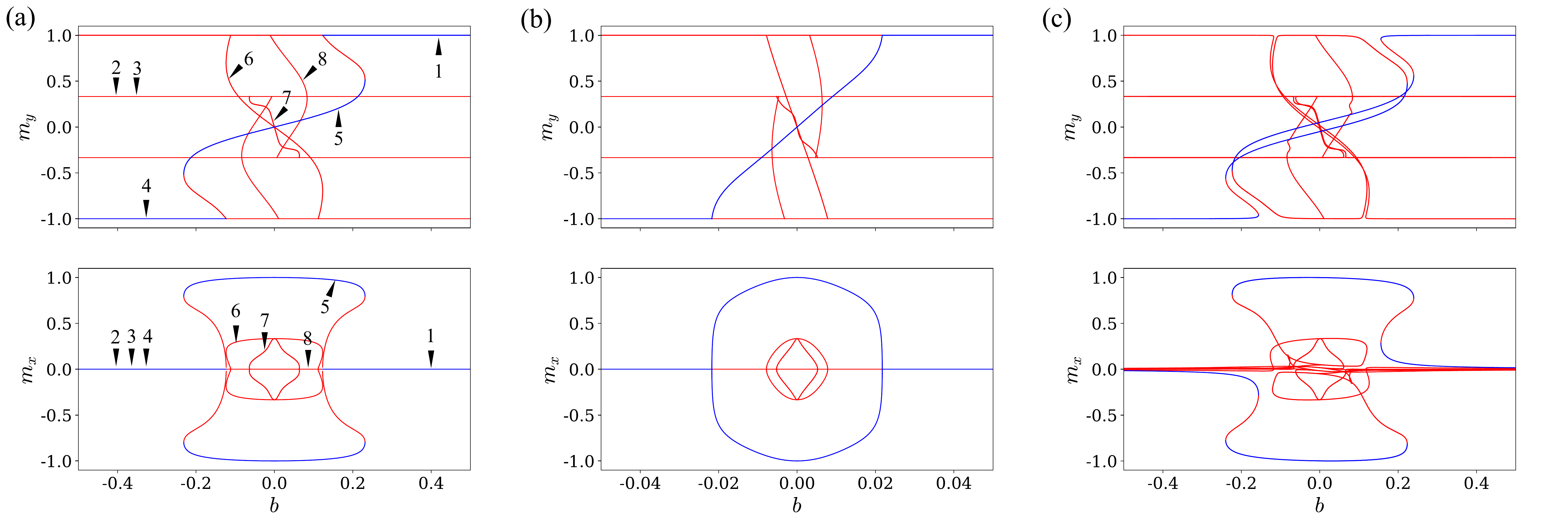}
\includegraphics[width=\textwidth]{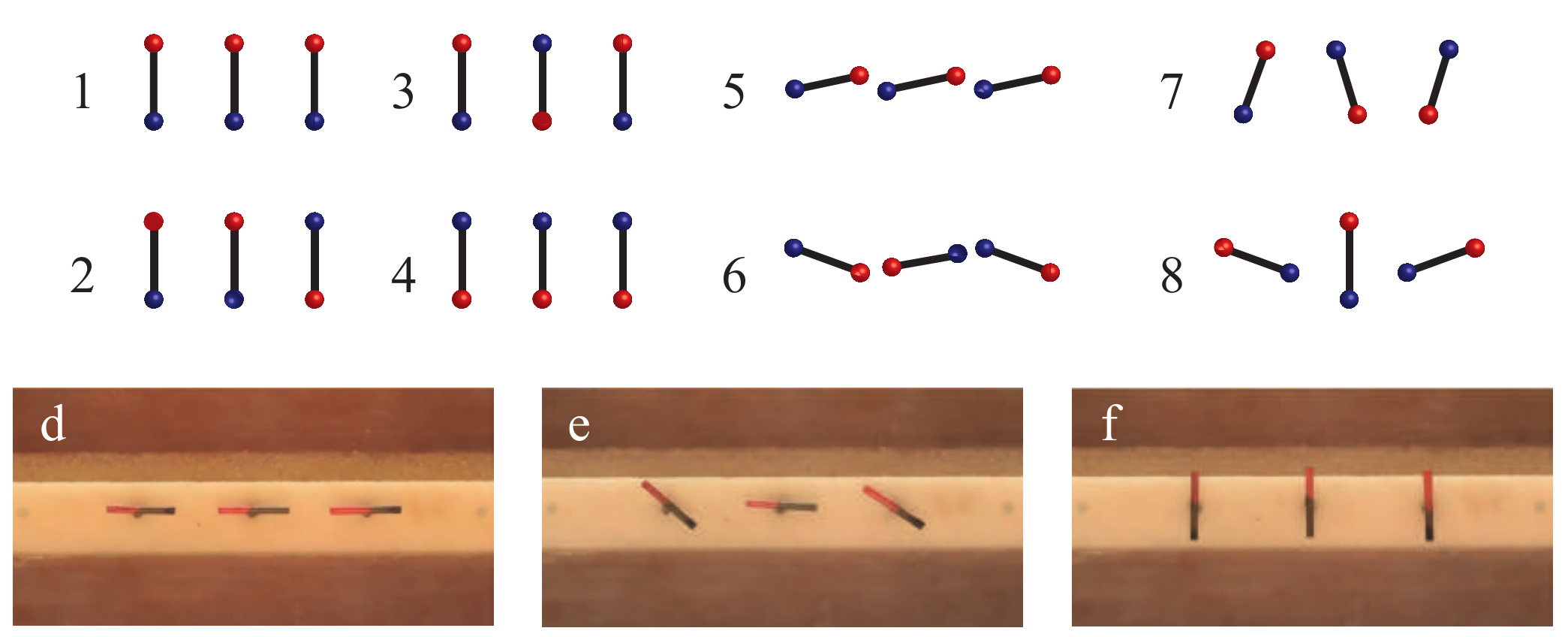}
\end{center}
\caption{(a) Bifurcation diagram for $N=3$ magnetic dipoles with separation parameter $\Delta/a=6/5$.
Branches are projected on the $(m_y,b)$ and the $(m_x,b)$ planes.
Red lines represent unstable states; blue lines represent stable states.
Horizontal lines in $m_y$ diagram represent trivial parallel solutions.
Only two of these branches are stable within a wide range of $b$: all-up, and all-down.
They become unstable at two separate pitchfork bifurcations, as depicted in diagram $(b,m_x)$.
Intermediate horizontal lines in $(b,m_y)$ represent several parallel states that overlap;
some of them show additional pitchfork bifurcations but no stable solutions are born.
Stable canted solutions emerge from the center of $(m_x,b)$ diagram and die in the aforementioned pitchfork bifurcations.
(b) Case  $N=3$ with $\Delta/a=5$. Here there is no stable coexistence between canted states and the parallel states since the bifurcation is supercritical.
(c) Similar to (a) but now linear chain is tilted at 10 degrees with respect to $x$ axis shown in Fig.\ref{fig:dumbbells}(a).
Some fine details of diagrams are transformed as pitchfork bifurcations become saddle-node bifurcations,
and parallel trivial states are not always solutions.
(d-f) Experimental stable configurations found for $N=3$. In the Supplementary information, a movie with the evolution of the states and the sudden transition from fully polarized to ferromagnetic is provided for clarity {\bf{Movie1SI}}. The numbered configurations  correspond to the different branches shown in (a).
}
\label{fig:diagram3}
\end{figure*}

\section{Stable and unstable solutions}

Using numerical continuation software AUTO \cite{doedel1997continuation} and considering all the different trivial solutions (parallel and aligned) at $b=0$ and at $b=\pm \infty$ as seed solutions,
we exhaustively computed a large number of branches of solutions of Eq.~(\ref{model}) for a wide interval
of values of control parameter $b$. There may, however, exist other special branches that live within limited regions of $b$ that do not include $b=0$; the study of these special branches will require ad-hoc exploration.

The characterization of the stability of the solutions can be performed using the eigenvalues of the $2N\times 2N$ Jacobian matrix of Eq.~(\ref{model}) at the equilibria:
stable solutions have eigenvalues with negative real parts only. Stable solutions correspond to minima of the potential energy of the system.
In the $(b,m_y)$  and the $(b,m_x)$ diagrams we show all the branches that were found for a given selection of $N$ and $\Delta/a$.
The sections of stable solutions will be highlighted (by {\color{blue}\bf{blue lines}}) in these diagrams as they are the ones that can be observed in adiabatic experiments.

We can recognize the following special points in the bifurcation diagrams:
turning points and pitchfork bifurcations, points where a branch makes a turning point or fold. 
If the solution switches stability at the turning point, it is a saddle-node bifurcation.
If the solution remains unstable before and after the turning point, it is a saddle-saddle bifurcation.

Points of a branch where another branch is born are associated to the presence of symmetries in the solutions.
The symmetry of the branch that is born is usually lower (its isotropy subgroup has a smaller number of transformations than the parent branch). A change in stability may occur sometimes at these points.
\begin{figure*}
\begin{center}
\includegraphics[width=\textwidth]{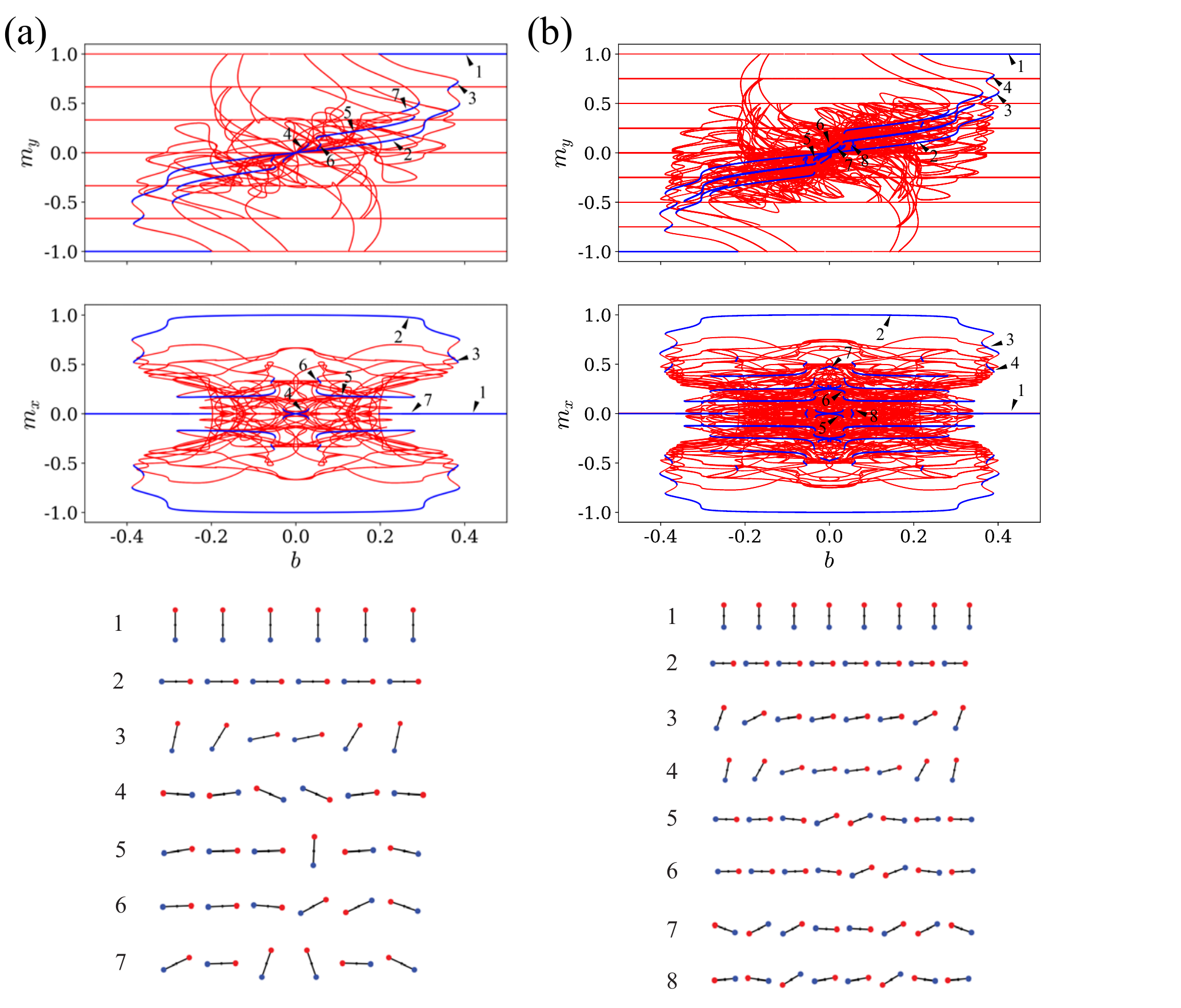}
\end{center}
\caption{(a) Bifurcation diagrams for $N=6$ magnetic dipoles with separation parameter $\Delta=6a/5$.
Red lines represent unstable states; blue lines represent stable states.
The branch that results from the aligned trivial state has additional saddle-node bifurcations
and new stable sections emerge. Configurations of the $N=6$ dipoles corresponding to numbered points are shown in the bottom panel.
(b) Bifurcation diagrams for $N=8$ magnetic dipoles with separation parameter $\Delta=6a/5$. Configurations of the $N=8$ dipoles corresponding to numbered points are shown in the bottom panel.
Both $N=6$ and $8$ cases show stable localized structures, even for $b=0$.
Compare with Fig. \ref{fig:dumbbells}(c-m).
}
\label{fig:diagram6}
\end{figure*}

We analyze in detail linear chains of size $N=3,6$, and $8$, because the first one can be fully described due to the small number of trajectories. On the other hand, $N=6$, and $N=8$ systems are great examples to show how the complexity of the system grows, but keeping the basic ideas of the $N=3$ untouched.
\begin{figure*}[t]
\begin{center}
\includegraphics[width=0.95\textwidth]{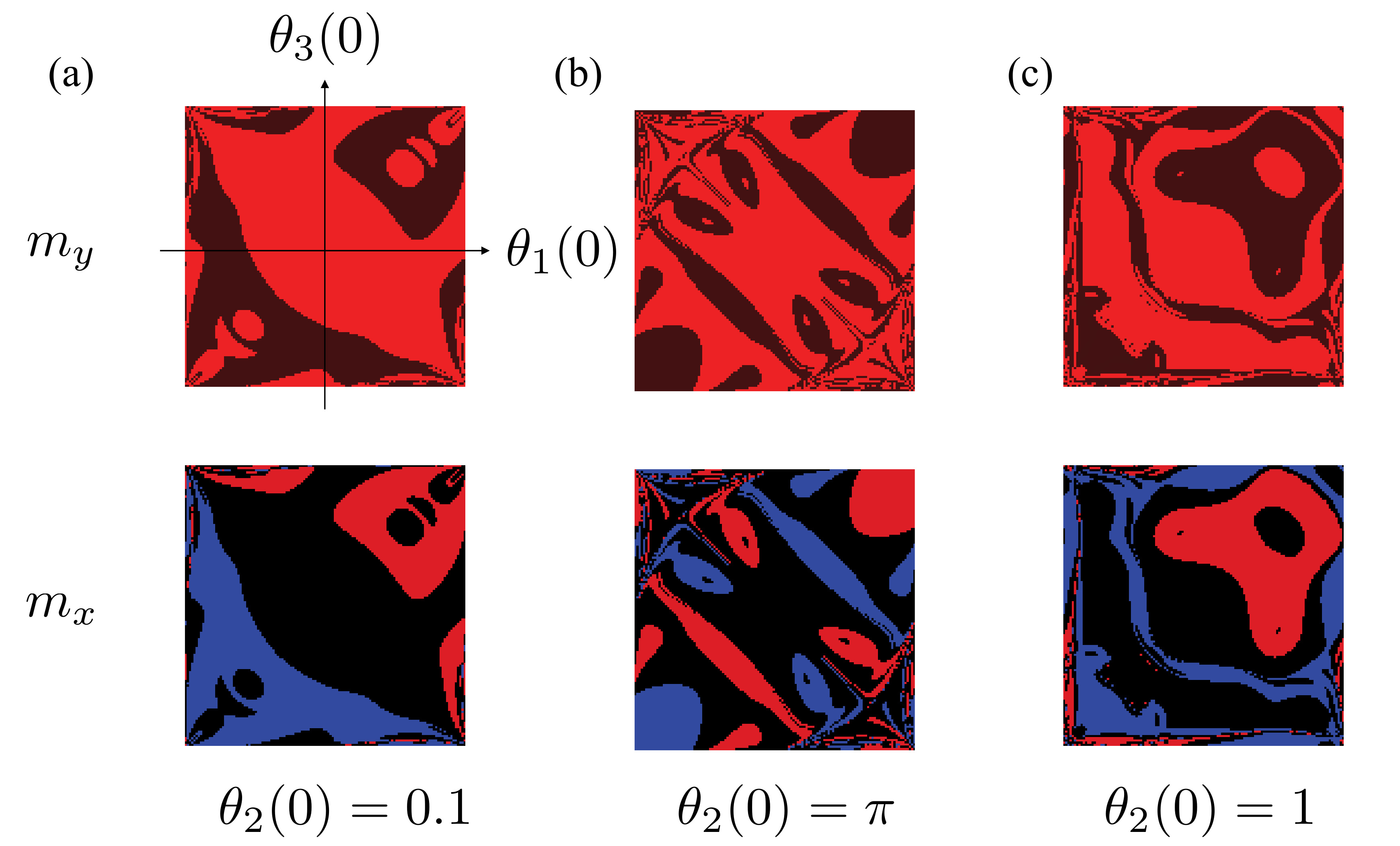}
\bigskip
\end{center}
\caption{Basins of attraction for $N=3$ dipoles featuring 3 stable equilibria.
Other parameters: $b=0.2, \Delta/a=6/5, \eta/I=0.133, \dot{\theta}_\alpha(0)=0$.
Each column corresponds to a different choice of $\theta_2(0)$,
selecting a 2D section in the 6D space of initial configurations.
Diagrams show $m_y$ and $m_x$ of the final state after a sufficiently long time. 
Red points indicate positive values; blue points indicate negative values.
Plots in the first row show vertical magnetization $m_y$ 
and plots in the second row show horizontal magnetization $m_x$.
(a) $\theta_2(0)=0.1$: all-up state ($m_y=1, m_x=0$) is dominant, canted right and canted left are unequally represented.
(b) $\theta_2(0)=\pi$: preserve balance between left and right states.
(c) $\theta_2(0)=1$: similar as in (a).
}
\label{fig:basins3}
\end{figure*}

For $N=3$ and $\Delta/a \sim 1$, shown in Fig.~\ref{fig:diagram3}(a),
the $(b,m_y)$ diagram shows several horizontal lines that correspond to trivial parallel solutions.
In $(b,m_x)$ diagram all these branches coincide in a single horizontal line $m_x=0$.
The all-up solution corresponds to $m_y=1$ and has a stable section beginning at some
positive value of $b$. A similar observation can be made for the all-down solution.
Their isotropy subgroups are both $\Sigma_\theta = \Ztwo(\nu) \times \Ztwo(\kappa)$.

Several canted states emerge from trivial aligned states $\theta_\alpha = \pm \pi/2$ at $b=0$.
They are unstable with the exception of the state with all dipoles canted in roughly the same direction,
associated to a branch with a `reversed S' shape that features two saddle-node bifurcations
and connects with all-up and all-down at two pitchfork bifurcations (as can be appreciated from the $(b,m_x)$ diagram).

The stable canted state (label 5 in Fig.~\ref{fig:diagram3}(a)) has the symmetry $\theta_1=\theta_3>\theta_2$ (an observation that applies to larger $N$).
Its isotropy subgroup is $\Sigma_\theta = \Ztwo(\kappa \nu)$.
There are two versions of this state, one canted to the right and another one to the left (that belong to the same group orbit).
For this particular choice of $\Delta$, this main canted branch makes a turning point and becomes unstable before
merging with the all-up and all-down branches. This is a \emph{sub-critical pitchfork}:
in a \emph{Gedanken} experiment, as the parameter $b$ is varied outside the stability region, the states of the dipoles rapidly rearrange and relax to either the all-up or the all-down states.

Branches beginning at states featuring other combinations of $\pm \pi/2$
are unstable and touch states $\theta=0, \pi$ after some saddle-saddle bifurcations
that do not involve changes in stability.
There is another canted state (label 6) that connects all-up and all-down but it is completely unstable.
It also has isotropy subgroup $\Ztwo(\kappa \nu)$, but their angles have different signs.
In contrast, the shorter branch (label 7 in Fig.~\ref{fig:diagram3}(a)) that connects up-down-down with down-up-up does not exhibit any symmetry.
\begin{figure*}
\begin{center}
\includegraphics[width=0.75\textwidth]{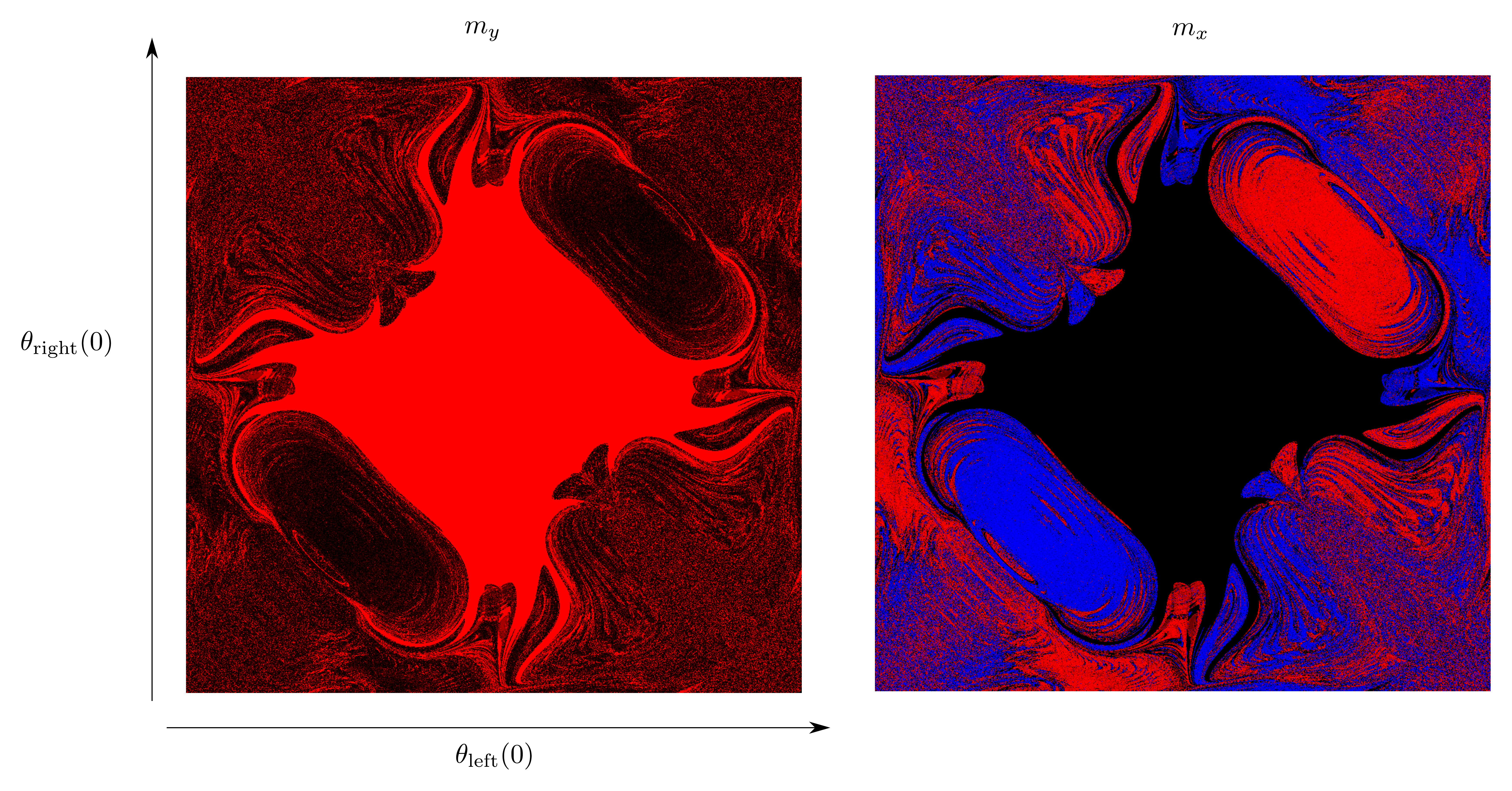}
\bigskip
\end{center}
\caption{Diagrams showing $m_y$ and $m_x$ of the final state after a sufficiently long time. They are the basins of attraction for a system of $N=6$ dipoles featuring 3 stable equilibria.
In this figure the parameters are: $b=0.2, \Delta=6a/5, \eta/I=0.133, \dot{\theta}_\alpha(0)=0$. 
Red points indicate positive values; blue points indicate negative values.
Given the enormous (12D) space of initial configurations we have used the following 2D section:
$\theta_1(0)=\theta_2(0)=\theta_3(0)=\theta_\mathrm{left}(0)$ and
$\theta_4(0)=\theta_5(0)=\theta_6(0)=\theta_\mathrm{right}(0)$.
The all-up state ($m_y=1, m_x=0$) is dominant, but there are large sections of the maps that are extremely sensitive 
to initial conditions.
 }
\label{fig:basins6}
\end{figure*}

There are two branches associated to `cross' solutions (label 8) that feature the central dipole pointing always in the same direction, while the other two perform a symmetric motion: $\theta_2=0$ (or $\pi$), $\theta_1=-\theta_3$.
These peculiar solutions are always unstable and only exist for odd $N$.
Their isotropy subgroups are $\Ztwo(\nu)$, connecting all-up with down-up-down, or all-down with up-down-up.

These diagrams show two tristable regions, agreeing with experiments \cite{concha2018designing}, where ferromagnetic, canted, and fully polarized states were observed for different magnetic fields (See Fig.\ref{fig:diagram3} (d-f) ).
The whole $(b,m_y)$ diagram is symmetric under the reflection through the center because of $\Ztwo(\mu)$.
The whole $(b,m_x)$ diagram is symmetric under the reflection with respect to the horizontal axis because of $\Ztwo(\kappa)$,
symmetric under the reflection with respect to vertical axis because of $\Ztwo(\mu)$.

For larger separation between dipoles, for instance $\Delta=5a$, most of the features of
the diagrams remain but there are some differences that we emphasize.
In Fig.~\ref{fig:diagram3}(b) we show the corresponding diagrams.
Canted states live in a narrower interval of parameter $b$,
their branches are now more straight, and there is no coexistence between stable canted states
and stable all-up or all-down states. The pitchfork bifurcations are \emph{super-critical}
since the dipoles smoothly become vertical as $b$ is varied. 
The diagrams have the same symmetries as noted for smaller $\Delta/a$.

These findings suggest that by using $b$ and $\Delta/a$ as control parameters,
not only the bifurcations are going to shift, but whole new stable branches are going to be created
by pairs of folds along certain branches.

Some of the symmetries can be removed to unfold 
the degeneracies of the $(b,m_y)$ and $(b,m_x)$ diagrams.
In Fig.~\ref{fig:diagram3}(c), the linear array of dipoles is oriented not perpendicular to the direction of $\vec{B}$
but subtending an anglñe of 80 degrees.
The model Eq.~(\ref{model}) is still valid, but the locations of the centers of rotation of the dipoles $(x_\alpha, y_\alpha)$ are now different.
As a result, the symmetry $\Ztwo(\kappa)$ is now broken.
Parallel trivial states $\theta=0, \pi$ are no longer solutions for finite $b$
and the canted states to the right and to the left unfold into two separate curves.
Pitchfork bifurcations become saddle-nodes.
Interestingly, both $(b,m_y)$ and $(b,m_x)$ diagrams are still symmetric under reflection through the center because of $\Ztwo(\mu)$.

For larger $N$, the number of branches grows exponentially as a result of the
large number of combinations that define trivial states.
The number of bifurcations also grows, and for small $\Delta/a$, the shape of the curves
associated to canted solutions develops many turns.
All these phenomena facilitate the existence of multiple stable states that we observe in experiments. See Fig. \ref{fig:dumbbells}(d-f) and Fig. \ref{fig:dumbbells}(j-m) for $N=11$ and $N=8$ respectively. 

For instance, for $N=6$ and $\Delta/a=6/5$ depicted in Fig.~\ref{fig:diagram6}(a),
the number of red curves (unstable) is quite large, but the messiness is only
superficial since both diagrams are still organized by the same symmetries.
The main stable canted branch becomes `wavy' featuring more saddle-node bifurcations and new stable sections.
These small stable sections are relevant because they explain small jumps in magnetization
that correspond to sudden jumps in the extreme dipoles that become more vertical
as $b$ is varied beyond the saddle-nodes.

There are other branches that acquire stable sections after one or more folds. There are also branches with short stable sections not limited by turning points: these changes in stability generate additional branches of reduced symmetry that may have stable solutions. All these symmetry-breaking bifurcations follow the predictions of the isotropy lattice shown in Fig. 2.
The new stable states are interesting because the dipoles are oriented in different directions, and depending on the magnetic texture, may form new objects that can be manipulated to store or process information \cite{parkin2008magnetic}. We have confirmed the existence of some of the numerically found states. These highly nontrivial heterogeneous stable solutions are shown in Fig. \ref{fig:dumbbells} (c-m).


For $N=8$ and small $\Delta=6a/5$, as depicted in Fig.~\ref{fig:diagram6}(b),
we found a more extreme growth in the number of unstable solutions.
But the number of new stable states is even more remarkable.
There is even a new state for $b=0$, that could be observed without an external magnetic field. The state $5$ in Fig. \ref{fig:diagram6}(b) was experimentally observed and a double defect is shown in Fig. \ref{fig:dumbbells}(j) similar to the predicted state $7$ in Fig. \ref{fig:diagram6}(b). These solutions should be `programmable' by direct manipulation of the dipoles, opening an opportunity for 
the development of XY-metamaterials. The structure of these new stable solutions is reminiscent of `magnetic domains'.
This is similar to localized structures (a few vertical dipoles surrounded by mostly horizontal ones) and `snaking' (sequences of saddle-nodes and saddle-saddle bifurcations) in the context of continuous 1D media
(see for instance the review \cite{knobloch2015spatial}).


\section{Basins of attraction}

In the previous section we have unveiled branches of solutions of the dipole equations
that can be traced to some of the trivial (parallel or aligned) solutions. Although most of these new nontrivial branches
are unstable, there are some sections that are stable and thus can be found in experiments (See Fig.~\ref{fig:dumbbells}(c-m))). Now for the characterization of stability, we have used the real part of the eigenvalues of the Jacobian of the system.
\begin{figure*}[t]
\begin{center}
\includegraphics[width=\textwidth]{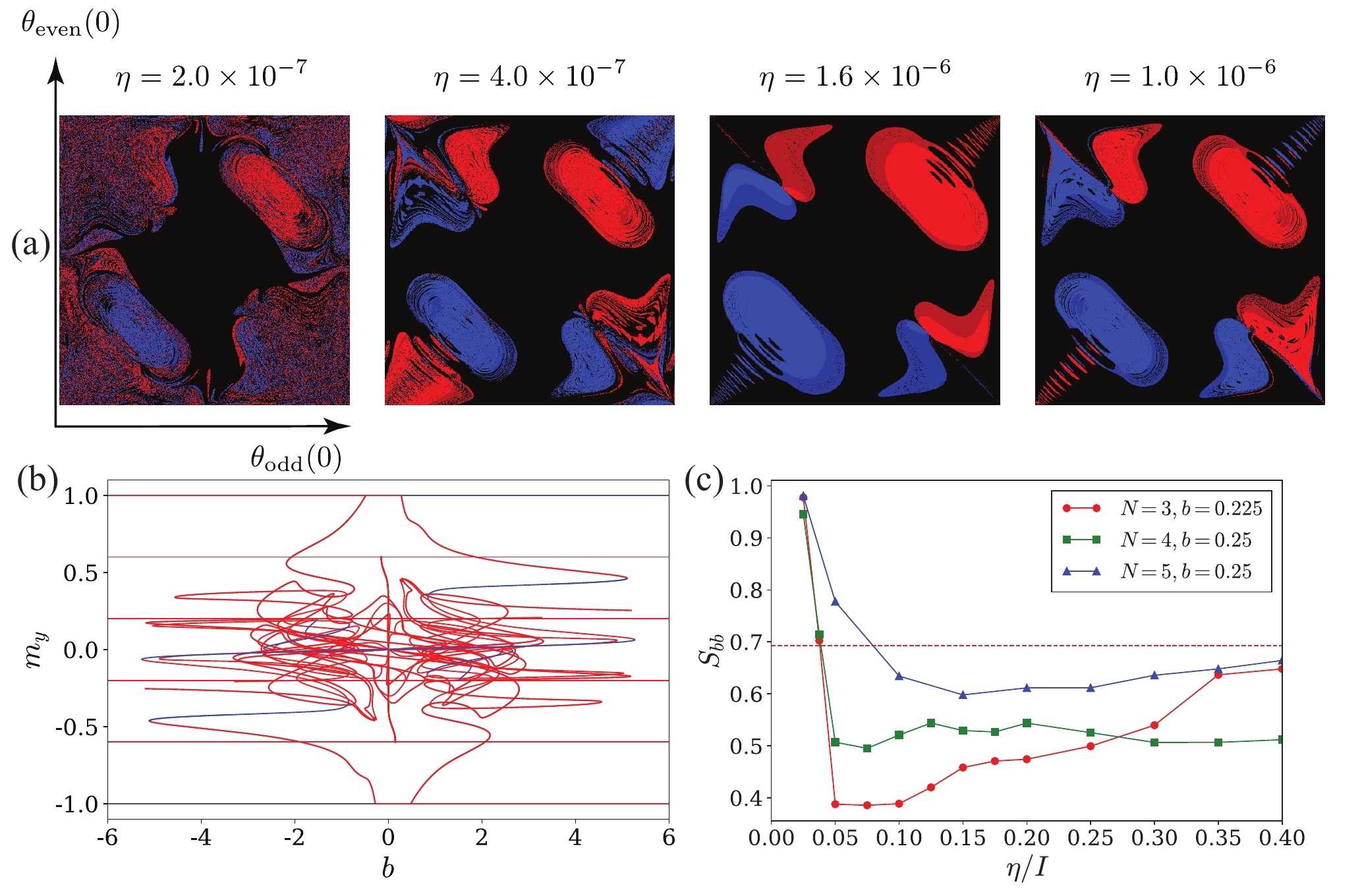}
\medskip
\end{center}
\caption{Analysis of basins of attraction for $N=5$ dipoles featuring 5 stable equilibria.
Other parameters: $b=1, \Delta/a=0.2, \eta/I=0.133, \dot{\theta}_\alpha(0)=0$.
(a) Diagrams show $m_x$ of the final state after a sufficiently long time for four selected values of $\eta$.
Red points indicate positive values; blue points indicate negative values.
The following 2D section is used: $\theta_1(0)=\theta_3(0)=\theta_5(0)=\theta_\mathrm{odd}(0)$ (horizontal coordinate),
and $\theta_2(0)=\theta_4(0)=\theta_\mathrm{even}(0)$ (vertical coordinate). 
(b) Bifurcation diagram $(b,m_y)$ showing coexistence of several stable branches, in particular for $b=1$ there are 5 stable states.
(c) Boundary basin entropy $S_{bb}$ as a function of friction $\eta$ and for three different numbers of dipoles,
showing a monotonic behavior, and that for $\eta>2.5 \times 10^{-7}$, the entropy $S_{bb}<\log 2$ suggests that the boundaries are fractal.
}
\label{fig:basins5}
\end{figure*}
The criterion of eigenvalues is not the only characterization or the most effective one.
Another criterion is the analysis of the potential energy associated to the torque equations:
\begin{equation}
U =  \sum_{\alpha, \beta} \sum_{i \in \alpha, j \in \beta}
\frac{\mu_0}{4 \pi} \frac{Q_i Q_j}{r_{ij}} - \sum_\alpha \sum_{i \in \alpha} a_i |Q_i| B \cos \theta_\alpha 
\end{equation}

The depths and widths of the energy minima give information about the attractiveness of each
stable equilibria, in particular in the presence of random perturbations that generate transitions between minima (as considered for instance in Ref. \cite{plihon2016stochastic}).


In the context of arbitrary initial conditions for the angles,
an interesting idea is the concept of basins of attraction \cite{nusse1996basins,grebogi1986multi}:
a map from the $2N$-dimensional space of initial configurations $\theta_\alpha(0), \dot{\theta}_\alpha(0)$
to the stable states that are reached after a long time.
Now, in contrast to the static case, inertia and friction are relevant. This is not only because they determine the time scale of the relaxation dynamics, but also because they determine the details of the trajectories and the final energy minimum that is reached.
The ratios of the volumes of the basins of attraction indicate the relevance of the different stable static solutions
(see \cite{menck2013basin} for a different approach).

As the dimension of the space of initial configurations
$\theta_\alpha(0), \dot{\theta}_\alpha(0)$ is large, 
we should explore 2D sections, for instance by fixing $\dot{\theta}_\alpha(0)$
or by imposing restrictions on $\theta_\alpha(0)$ and $\theta_\beta(0)$.

Setting specific values of $N, b, \Delta$ that guarantee coexistence of two or more stable states, we have detected
boundaries of basins of attraction that show unexpected richness and beauty.
Even if the dynamics is relatively simple, the basins are not.

For instance, for $N=3$ and three stable states, as depicted in Fig.~\ref{fig:basins3}:
a parallel trivial state with $m_y=1, m_x=0$;
and two canted states with $m_y<1$ and $m_x$ of equal magnitude and opposite sign.
We found interesting shapes for the three basins of attraction,
by using a value of $b$ such that the three basins have similar volumes.
Using different values $\theta_2(0)$ and setting $\dot{\theta}_\alpha = 0$,
we found 2D sections with unexpected shapes.

The map of basins will inherit the symmetries of the selected section.
For instance, by choosing $\theta_2(0) = 0$ or $\pi$,
points $(\theta_1(0),\theta_3(0))$, $(\theta_3(0),\theta_1(0))$ and
$(-\theta_1(0),-\theta_3(0))$, and $(-\theta_3(0),-\theta_1(0))$
belong to the same group orbit, and the 2D sections will have symmetry under reflection with respect to the diagonal $\theta_1(0)=\theta_3(0)$
and under reflection with respect to the diagonal $\theta_1(0)=-\theta_3(0)$
(as shown in the central column of Fig. \ref{fig:basins3}).
On the other hand, by choosing $\theta_2(0) \ne 0$ or $\pi$,
$(\theta_1(0),\theta_3(0))$ and $(-\theta_1(0),-\theta_3(0))$ no longer will be in the same group orbit,
and the 2D section will only be symmetric under reflection with respect to the diagonal $\theta_1(0)=\theta_3(0)$
(as shown in left and right columns of Fig. \ref{fig:basins3}).

For more dipoles and smaller damping, the boundaries of the basins can become rugged and even fractal.
As the volume occupied by these fractal structures grows, the dynamics becomes extremely sensitive to
changes in initial configurations.
It is possible to measure the capacity dimension of the boundary that separate the basins and make
a connection with an exponent that measures the uncertainty \cite{grebogi1983final}.
For $N=6$ and three stable states, see Fig.~\ref{fig:basins6}, similar to the previous case,
the 12-dimensional space of initial configurations can be explored by
suitable 2-dimensional sections.
One possibility is setting $\dot{\theta}_\alpha(0)=0$ as before and now locking $\theta_1(0)=\theta_2(0)=\theta_3(0)$
and $\theta_4(0)=\theta_5(0)=\theta_6(0)$.
Both diagrams are symmetric under reflections with respect to both diagonals.

The all-up state ($m_y=1, m_x=0$) is dominant but there are large sections of the maps that show
extreme sensitivity to initial conditions: transiently chaotic dynamics in the presence of dissipation.
There are two billiard-shape regions, one with positive $m_x$ and other one with negative,
as well as `arms' from the all-up state into the chaotic regions.
\begin{figure*}[t]
\begin{center}
\includegraphics[width=0.95\textwidth]{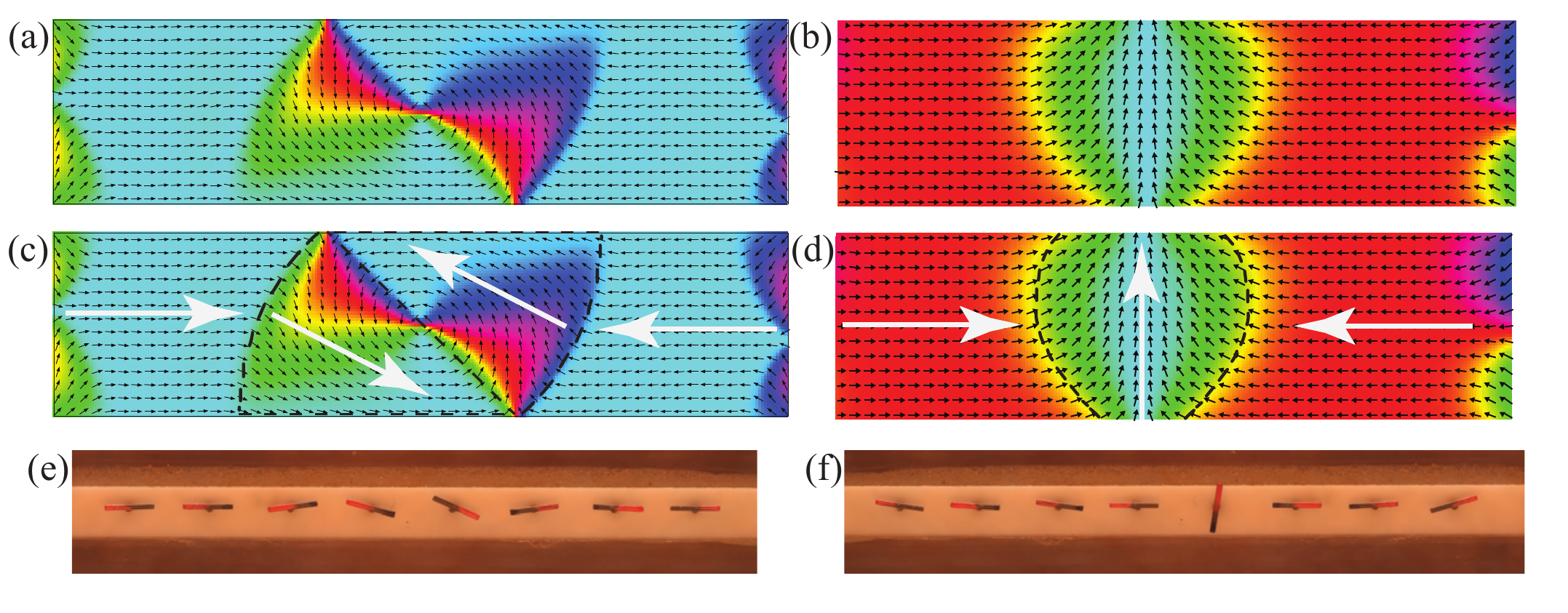}
\medskip
\end{center}
\caption{Comparison of the domain walls found in permalloy thin films \cite{mcmichael1997head,clarke2008dynamics}, and in a dipolar chain. In  ({\bf{a,b}}) micromagnetic simulations performed with the OOMMF software are shown \cite{donahue1999oommf}. The simulation parameters for ({\bf{a}}) are $L=1000$nm, $W=250$ nm, $t=32$ nm and material parameters typical for permalloy, and for ({\bf{b}}) $L=1000$nm, $W=250$ nm, $t=2$ nm. In ({\bf{c-d}}) we show coarse grained interpretations of the numerical simulations that consider four regions and their average magnetization represented by white arrows. ({\bf{e-f}}) are the solutions experimentally found in our macroscopic experiment for $N=8$, and that can also be observed in Fig.\ref{fig:dumbbells}. 
}
\label{fig:domains}
\end{figure*}

The sensitivity to changes in initial conditions can be quantified using the concept
of \emph{boundary basin entropy} developed by Daza et al. \cite{daza2016basin}
from a discretized map:
\begin{equation}
S_{bb} = \frac{1}{N N_b} \sum^{N}_{i=1} \sum^{m_i}_{j=1} p_{i,j} \log \left( \frac{1}{p_{i,j}}\right) \hspace{1cm},
\end{equation}
where $i=1\ldots N$ labels all the boxes of size $\varepsilon$;
$j=1\ldots m_i$ labels the possible states reached by initial conditions in that particular box;
$p_{i,j}$ is the probability that an initial condition inside the box reaches state $j$.
$N_b$ is the number of boxes that has more than one final state.
It quantifies the roughness of the boundaries: the uncertainty referring only to the boundaries,
without taking into account the volumes of the basins.
It provides a sufficient condition to assess that some boundaries are fractal. Fragmented boundaries are indication of diverging trajectories that spend a long time exploring different regions of phase space (a phenomena known as transient chaos) before reaching a stable configuration.
Smooth boundaries possess low $S_{bb}$; rugged boundaries, high $S_{bb}$.
More specifically, for the boundary to be fractal
a sufficient but not necessary condition is $S_{bb} > \log 2$  \cite{daza2016basin}.

Here we do not assess the entropy of the whole 2N dimensional space but of selected 2D sections. For instance $N=3$ in Fig.~\ref{fig:basins3}, the values of $S_{bb}$ are $0.5698$, $0.5803$ and $0.5936$ respectively. 
For $N=6$ in Fig.~\ref{fig:basins6}, $S_{bb}=0.8056$, that indicates fractal boundaries.

As an example, while analyzing the case $N=5$ for several values of the friction coefficient $\eta$,
we found fractal boundaries of basins. Increasing the friction the boundaries
become smoother and $S_{bb}$ decreases below $\log2$.
In Fig.~\ref{fig:basins5} (a), we show the impact of the friction coefficient.
We use $\theta_\mathrm{odd}(0)=\theta_1(0)=\theta_3(0)=\theta_5(0)$ and $\theta_\mathrm{even}(0)=\theta_2(0)=\theta_4(0)$.

Since $(\theta_\mathrm{odd}(0),\theta_\mathrm{even}(0))$ and $(-\theta_\mathrm{odd}(0),-\theta_\mathrm{even}(0))$
belong to the same group orbit
the whole diagram will be symmetric under reflection through the center point. 

The entropy was also computed for $N=3$, $N=4$, and $N=5$ for several values of $\eta/I$, allowing us to verify the intuitive idea 
that at low damping the basins become fractal. However, no estimate for the transition value is provided. 

\section{Micromagnetic simulations}
To better understand the relevance of the localized solutions predicted in diagrams such as Fig.(\ref{fig:diagram6}), and experimentally found Fig.(\ref{fig:domains}), we note that they are extremely similar to the well known Neel and Bloch DW solutions first reported by McMichael and Donahue \cite{mcmichael1997head} in the context of micromagnetic simulations \cite{donahue1999oommf}, and later explained in terms of topological charges \cite{tchernyshyov2005fractional}. To make this apparent, we draw the coarse grained versions of the domains  found in the micromagnetic simulations, and compare these textures with the magnetic bars used in our experiments. The vector field generated by the reported DW has a texture that is topologically equivalent to the one seen in permalloy thin films. This clear analogy opens a door for further miniaturization of domain wall logic, as our results show that dipolar systems similar to the classical realization shown here can display localized structures. Therefore, racetrack memories or related devices \cite{parkin2008magnetic} are likely to be miniaturized down to the scale of the smaller dipolar objects \cite{cambre2015asymmetric}.   
 
\section{Conclusions}
In this article we have studied a simple, yet extremely rich system of polar objects that are endowed with XY local symmetry, and are arranged in space along a line. Despite its apparent simplicity, interactions and symmetries conspire to produce the fine structure of the magnetic response of the system which is a reflection of the symmetries of the system, and therefore depends on the full set of possible solutions, on whether they are stable or unstable. The numerical and experimental results have shown that there are Neel and Bloch types of defects that are stable in large regions of the parameter space. 

Furthermore, for large enough chains of polar objects many defects are supported and there are possible stable solutions. The collision of two domain walls allows the transition from one solution to another. Typically this transition is accompanied by the emission of spin waves. However, we believe that by a resonant driving, it should be possible to pump DW in a system of this type. 

Overall, the richness of the possible solutions found in this system allows us to describe  all the details of the hysteresis curves found in \cite{concha2018designing}, and highlights the role of group theory to describe the possible solutions in this type of systems. Moreover, we have shown that these DWs are stable in a large region of the parameters space, even when chains have very few components, therefore this is a call to push for the miniaturization of DW logic using race track memories \cite{parkin2008magnetic}.

We close by pointing out that the analysis performed in this system (the description of all possible solutions) allows for the easy computation of the probability amplitude by summing over all (stable and unstable) histories {\it{\`a la}} Feynman. Thus, defining a path integral over all classical solutions, regardless of its stability, will provide an extremely useful tool for the computation of interference effects in magnetic clusters. In the case of classical systems, the unstable trajectories cannot be measured with an adiabatic experiment, and they are hard to measure even in parametrically forced situations. However, as the {\it{quantumness}} of the system increases those trajectories define the fate of the system. This observation makes clear that the characterization of all the classical trajectories and the symmetries that enforce transitions in real space are a key factor for our understanding of atomic magnetic clusters \cite{PhysRevLett.125.237602}.

\section*{Acknowledgments}
F.U. thanks FONDECYT (Chile) for financial support through Postdoctoral  3180227.
C.P acknowledge support from the CM-iLab.
A.C. acknowledge support from the CODEV Seed Money Programme of the \'Ecole Polytechnique F\'ed\'erale de Lausanne (EPFL), and the support of the Design Engineering Center at UAI. 


\begin{thebibliography}{46}%
\makeatletter
\providecommand \@ifxundefined [1]{%
 \@ifx{#1\undefined}
}%
\providecommand \@ifnum [1]{%
 \ifnum #1\expandafter \@firstoftwo
 \else \expandafter \@secondoftwo
 \fi
}%
\providecommand \@ifx [1]{%
 \ifx #1\expandafter \@firstoftwo
 \else \expandafter \@secondoftwo
 \fi
}%
\providecommand \natexlab [1]{#1}%
\providecommand \enquote  [1]{``#1''}%
\providecommand \bibnamefont  [1]{#1}%
\providecommand \bibfnamefont [1]{#1}%
\providecommand \citenamefont [1]{#1}%
\providecommand \href@noop [0]{\@secondoftwo}%
\providecommand \href [0]{\begingroup \@sanitize@url \@href}%
\providecommand \@href[1]{\@@startlink{#1}\@@href}%
\providecommand \@@href[1]{\endgroup#1\@@endlink}%
\providecommand \@sanitize@url [0]{\catcode `\\12\catcode `\$12\catcode
  `\&12\catcode `\#12\catcode `\^12\catcode `\_12\catcode `\%12\relax}%
\providecommand \@@startlink[1]{}%
\providecommand \@@endlink[0]{}%
\providecommand \url  [0]{\begingroup\@sanitize@url \@url }%
\providecommand \@url [1]{\endgroup\@href {#1}{\urlprefix }}%
\providecommand \urlprefix  [0]{URL }%
\providecommand \Eprint [0]{\href }%
\providecommand \doibase [0]{http://dx.doi.org/}%
\providecommand \selectlanguage [0]{\@gobble}%
\providecommand \bibinfo  [0]{\@secondoftwo}%
\providecommand \bibfield  [0]{\@secondoftwo}%
\providecommand \translation [1]{[#1]}%
\providecommand \BibitemOpen [0]{}%
\providecommand \bibitemStop [0]{}%
\providecommand \bibitemNoStop [0]{.\EOS\space}%
\providecommand \EOS [0]{\spacefactor3000\relax}%
\providecommand \BibitemShut  [1]{\csname bibitem#1\endcsname}%
\let\auto@bib@innerbib\@empty
\bibitem [{\citenamefont {O'handley}(2000)}]{handley2000modern}%
  \BibitemOpen
  \bibfield  {author} {\bibinfo {author} {\bibfnamefont {R.~C.}\ \bibnamefont
  {O'handley}},\ }\href@noop {} {\emph {\bibinfo {title} {Modern magnetic
  materials: principles and applications}}},\ Vol.\ \bibinfo {volume}
  {830622677}\ (\bibinfo  {publisher} {Wiley New York},\ \bibinfo {year}
  {2000})\BibitemShut {NoStop}%
\bibitem [{\citenamefont {Coey}(2010)}]{coey2010magnetism}%
  \BibitemOpen
  \bibfield  {author} {\bibinfo {author} {\bibfnamefont {J.~M.}\ \bibnamefont
  {Coey}},\ }\href@noop {} {\emph {\bibinfo {title} {Magnetism and magnetic
  materials}}}\ (\bibinfo  {publisher} {Cambridge university press},\ \bibinfo
  {year} {2010})\BibitemShut {NoStop}%
\bibitem [{\citenamefont {Merz}(1953)}]{merz1953double}%
  \BibitemOpen
  \bibfield  {author} {\bibinfo {author} {\bibfnamefont {W.~J.}\ \bibnamefont
  {Merz}},\ }\href@noop {} {\bibfield  {journal} {\bibinfo  {journal} {Physical
  Review}\ }\textbf {\bibinfo {volume} {91}},\ \bibinfo {pages} {513} (\bibinfo
  {year} {1953})}\BibitemShut {NoStop}%
\bibitem [{\citenamefont {Woodward}\ and\ \citenamefont
  {DellaTorre}(1960)}]{woodward1960particle}%
  \BibitemOpen
  \bibfield  {author} {\bibinfo {author} {\bibfnamefont {J.}~\bibnamefont
  {Woodward}}\ and\ \bibinfo {author} {\bibfnamefont {E.}~\bibnamefont
  {DellaTorre}},\ }\href@noop {} {\bibfield  {journal} {\bibinfo  {journal}
  {Journal of Applied Physics}\ }\textbf {\bibinfo {volume} {31}},\ \bibinfo
  {pages} {56} (\bibinfo {year} {1960})}\BibitemShut {NoStop}%
\bibitem [{\citenamefont {Concha}\ \emph {et~al.}(2018)\citenamefont {Concha},
  \citenamefont {Aguayo},\ and\ \citenamefont {Mellado}}]{concha2018designing}%
  \BibitemOpen
  \bibfield  {author} {\bibinfo {author} {\bibfnamefont {A.}~\bibnamefont
  {Concha}}, \bibinfo {author} {\bibfnamefont {D.}~\bibnamefont {Aguayo}}, \
  and\ \bibinfo {author} {\bibfnamefont {P.}~\bibnamefont {Mellado}},\
  }\href@noop {} {\bibfield  {journal} {\bibinfo  {journal} {Physical review
  letters}\ }\textbf {\bibinfo {volume} {120}},\ \bibinfo {pages} {157202}
  (\bibinfo {year} {2018})}\BibitemShut {NoStop}%
\bibitem [{\citenamefont {Ising}(1925)}]{ising1925beitrag}%
  \BibitemOpen
  \bibfield  {author} {\bibinfo {author} {\bibfnamefont {E.}~\bibnamefont
  {Ising}},\ }\href@noop {} {\bibfield  {journal} {\bibinfo  {journal}
  {Zeitschrift f{\"u}r Physik}\ }\textbf {\bibinfo {volume} {31}},\ \bibinfo
  {pages} {253} (\bibinfo {year} {1925})}\BibitemShut {NoStop}%
\bibitem [{\citenamefont {Onsager}(1944)}]{onsager1944crystal}%
  \BibitemOpen
  \bibfield  {author} {\bibinfo {author} {\bibfnamefont {L.}~\bibnamefont
  {Onsager}},\ }\href@noop {} {\bibfield  {journal} {\bibinfo  {journal}
  {Physical Review}\ }\textbf {\bibinfo {volume} {65}},\ \bibinfo {pages} {117}
  (\bibinfo {year} {1944})}\BibitemShut {NoStop}%
\bibitem [{\citenamefont {Kaufman}(1949)}]{kaufman1949crystal}%
  \BibitemOpen
  \bibfield  {author} {\bibinfo {author} {\bibfnamefont {B.}~\bibnamefont
  {Kaufman}},\ }\href@noop {} {\bibfield  {journal} {\bibinfo  {journal}
  {Physical Review}\ }\textbf {\bibinfo {volume} {76}},\ \bibinfo {pages}
  {1232} (\bibinfo {year} {1949})}\BibitemShut {NoStop}%
\bibitem [{\citenamefont {Randeria}\ \emph {et~al.}(1985)\citenamefont
  {Randeria}, \citenamefont {Sethna},\ and\ \citenamefont
  {Palmer}}]{randeria1985low}%
  \BibitemOpen
  \bibfield  {author} {\bibinfo {author} {\bibfnamefont {M.}~\bibnamefont
  {Randeria}}, \bibinfo {author} {\bibfnamefont {J.~P.}\ \bibnamefont
  {Sethna}}, \ and\ \bibinfo {author} {\bibfnamefont {R.~G.}\ \bibnamefont
  {Palmer}},\ }\href@noop {} {\bibfield  {journal} {\bibinfo  {journal}
  {Physical review letters}\ }\textbf {\bibinfo {volume} {54}},\ \bibinfo
  {pages} {1321} (\bibinfo {year} {1985})}\BibitemShut {NoStop}%
\bibitem [{\citenamefont {Sethna}\ \emph {et~al.}(1993)\citenamefont {Sethna},
  \citenamefont {Dahmen}, \citenamefont {Kartha}, \citenamefont {Krumhansl},
  \citenamefont {Roberts},\ and\ \citenamefont {Shore}}]{sethna1993hysteresis}%
  \BibitemOpen
  \bibfield  {author} {\bibinfo {author} {\bibfnamefont {J.~P.}\ \bibnamefont
  {Sethna}}, \bibinfo {author} {\bibfnamefont {K.}~\bibnamefont {Dahmen}},
  \bibinfo {author} {\bibfnamefont {S.}~\bibnamefont {Kartha}}, \bibinfo
  {author} {\bibfnamefont {J.~A.}\ \bibnamefont {Krumhansl}}, \bibinfo {author}
  {\bibfnamefont {B.~W.}\ \bibnamefont {Roberts}}, \ and\ \bibinfo {author}
  {\bibfnamefont {J.~D.}\ \bibnamefont {Shore}},\ }\href@noop {} {\bibfield
  {journal} {\bibinfo  {journal} {Physical Review Letters}\ }\textbf {\bibinfo
  {volume} {70}},\ \bibinfo {pages} {3347} (\bibinfo {year}
  {1993})}\BibitemShut {NoStop}%
\bibitem [{\citenamefont {Jacobs}\ and\ \citenamefont
  {Luborsky}(1957)}]{jacobs1957magnetic}%
  \BibitemOpen
  \bibfield  {author} {\bibinfo {author} {\bibfnamefont {I.}~\bibnamefont
  {Jacobs}}\ and\ \bibinfo {author} {\bibfnamefont {F.}~\bibnamefont
  {Luborsky}},\ }\href@noop {} {\bibfield  {journal} {\bibinfo  {journal}
  {Journal of Applied Physics}\ }\textbf {\bibinfo {volume} {28}},\ \bibinfo
  {pages} {467} (\bibinfo {year} {1957})}\BibitemShut {NoStop}%
\bibitem [{\citenamefont {Jiles}\ and\ \citenamefont
  {Atherton}(1986)}]{jiles1986theory}%
  \BibitemOpen
  \bibfield  {author} {\bibinfo {author} {\bibfnamefont {D.}~\bibnamefont
  {Jiles}}\ and\ \bibinfo {author} {\bibfnamefont {D.}~\bibnamefont
  {Atherton}},\ }\href@noop {} {\bibfield  {journal} {\bibinfo  {journal}
  {Journal of magnetism and magnetic materials}\ }\textbf {\bibinfo {volume}
  {61}},\ \bibinfo {pages} {48} (\bibinfo {year} {1986})}\BibitemShut {NoStop}%
\bibitem [{\citenamefont {Krasnosel'skii}\ \emph {et~al.}(2012)\citenamefont
  {Krasnosel'skii}, \citenamefont {Niezgodka},\ and\ \citenamefont
  {Pokrovskii}}]{krasnosel2012systems}%
  \BibitemOpen
  \bibfield  {author} {\bibinfo {author} {\bibfnamefont {M.}~\bibnamefont
  {Krasnosel'skii}}, \bibinfo {author} {\bibfnamefont {M.}~\bibnamefont
  {Niezgodka}}, \ and\ \bibinfo {author} {\bibfnamefont {A.}~\bibnamefont
  {Pokrovskii}},\ }\href@noop {} {\emph {\bibinfo {title} {Systems with
  Hysteresis}}}\ (\bibinfo  {publisher} {Springer Berlin Heidelberg},\ \bibinfo
  {year} {2012})\BibitemShut {NoStop}%
\bibitem [{\citenamefont {Ramirez}\ \emph {et~al.}(1999)\citenamefont
  {Ramirez}, \citenamefont {Hayashi}, \citenamefont {Cava}, \citenamefont
  {Siddharthan},\ and\ \citenamefont {Shastry}}]{ramirez1999zero}%
  \BibitemOpen
  \bibfield  {author} {\bibinfo {author} {\bibfnamefont {A.~P.}\ \bibnamefont
  {Ramirez}}, \bibinfo {author} {\bibfnamefont {A.}~\bibnamefont {Hayashi}},
  \bibinfo {author} {\bibfnamefont {R.~J.}\ \bibnamefont {Cava}}, \bibinfo
  {author} {\bibfnamefont {R.}~\bibnamefont {Siddharthan}}, \ and\ \bibinfo
  {author} {\bibfnamefont {B.}~\bibnamefont {Shastry}},\ }\href@noop {}
  {\bibfield  {journal} {\bibinfo  {journal} {Nature}\ }\textbf {\bibinfo
  {volume} {399}},\ \bibinfo {pages} {333} (\bibinfo {year}
  {1999})}\BibitemShut {NoStop}%
\bibitem [{\citenamefont {Castelnovo}\ \emph {et~al.}(2008)\citenamefont
  {Castelnovo}, \citenamefont {Moessner},\ and\ \citenamefont
  {Sondhi}}]{castelnovo2008magnetic}%
  \BibitemOpen
  \bibfield  {author} {\bibinfo {author} {\bibfnamefont {C.}~\bibnamefont
  {Castelnovo}}, \bibinfo {author} {\bibfnamefont {R.}~\bibnamefont
  {Moessner}}, \ and\ \bibinfo {author} {\bibfnamefont {S.~L.}\ \bibnamefont
  {Sondhi}},\ }\href@noop {} {\bibfield  {journal} {\bibinfo  {journal}
  {Nature}\ }\textbf {\bibinfo {volume} {451}},\ \bibinfo {pages} {42}
  (\bibinfo {year} {2008})}\BibitemShut {NoStop}%
\bibitem [{\citenamefont {Domic}\ \emph {et~al.}(2011)\citenamefont {Domic},
  \citenamefont {Goles},\ and\ \citenamefont {Rica}}]{domic2011dynamics}%
  \BibitemOpen
  \bibfield  {author} {\bibinfo {author} {\bibfnamefont {N.~G.}\ \bibnamefont
  {Domic}}, \bibinfo {author} {\bibfnamefont {E.}~\bibnamefont {Goles}}, \ and\
  \bibinfo {author} {\bibfnamefont {S.}~\bibnamefont {Rica}},\ }\href@noop {}
  {\bibfield  {journal} {\bibinfo  {journal} {Physical Review E}\ }\textbf
  {\bibinfo {volume} {83}},\ \bibinfo {pages} {056111} (\bibinfo {year}
  {2011})}\BibitemShut {NoStop}%
\bibitem [{\citenamefont {Kamminga}\ \emph {et~al.}(2018)\citenamefont
  {Kamminga}, \citenamefont {Hidayat}, \citenamefont {Baas}, \citenamefont
  {Blake},\ and\ \citenamefont {Palstra}}]{kamminga2018out}%
  \BibitemOpen
  \bibfield  {author} {\bibinfo {author} {\bibfnamefont {M.~E.}\ \bibnamefont
  {Kamminga}}, \bibinfo {author} {\bibfnamefont {R.}~\bibnamefont {Hidayat}},
  \bibinfo {author} {\bibfnamefont {J.}~\bibnamefont {Baas}}, \bibinfo {author}
  {\bibfnamefont {G.~R.}\ \bibnamefont {Blake}}, \ and\ \bibinfo {author}
  {\bibfnamefont {T.~T.}\ \bibnamefont {Palstra}},\ }\href@noop {} {\bibfield
  {journal} {\bibinfo  {journal} {APL materials}\ }\textbf {\bibinfo {volume}
  {6}},\ \bibinfo {pages} {066106} (\bibinfo {year} {2018})}\BibitemShut
  {NoStop}%
\bibitem [{\citenamefont {Gardner}\ \emph {et~al.}(2010)\citenamefont
  {Gardner}, \citenamefont {Gingras},\ and\ \citenamefont
  {Greedan}}]{gardner2010magnetic}%
  \BibitemOpen
  \bibfield  {author} {\bibinfo {author} {\bibfnamefont {J.~S.}\ \bibnamefont
  {Gardner}}, \bibinfo {author} {\bibfnamefont {M.~J.}\ \bibnamefont
  {Gingras}}, \ and\ \bibinfo {author} {\bibfnamefont {J.~E.}\ \bibnamefont
  {Greedan}},\ }\href@noop {} {\bibfield  {journal} {\bibinfo  {journal}
  {Reviews of Modern Physics}\ }\textbf {\bibinfo {volume} {82}},\ \bibinfo
  {pages} {53} (\bibinfo {year} {2010})}\BibitemShut {NoStop}%
\bibitem [{\citenamefont {Nisoli}\ \emph {et~al.}(2013)\citenamefont {Nisoli},
  \citenamefont {Moessner},\ and\ \citenamefont
  {Schiffer}}]{nisoli2013colloquium}%
  \BibitemOpen
  \bibfield  {author} {\bibinfo {author} {\bibfnamefont {C.}~\bibnamefont
  {Nisoli}}, \bibinfo {author} {\bibfnamefont {R.}~\bibnamefont {Moessner}}, \
  and\ \bibinfo {author} {\bibfnamefont {P.}~\bibnamefont {Schiffer}},\
  }\href@noop {} {\bibfield  {journal} {\bibinfo  {journal} {Reviews of Modern
  Physics}\ }\textbf {\bibinfo {volume} {85}},\ \bibinfo {pages} {1473}
  (\bibinfo {year} {2013})}\BibitemShut {NoStop}%
\bibitem [{\citenamefont {Pupillo}\ \emph {et~al.}(2009)\citenamefont
  {Pupillo}, \citenamefont {Micheli}, \citenamefont {B{\"u}chler},\ and\
  \citenamefont {Zoller}}]{pupillo2009condensed}%
  \BibitemOpen
  \bibfield  {author} {\bibinfo {author} {\bibfnamefont {G.}~\bibnamefont
  {Pupillo}}, \bibinfo {author} {\bibfnamefont {A.}~\bibnamefont {Micheli}},
  \bibinfo {author} {\bibfnamefont {H.-P.}\ \bibnamefont {B{\"u}chler}}, \ and\
  \bibinfo {author} {\bibfnamefont {P.}~\bibnamefont {Zoller}},\ }in\
  \href@noop {} {\emph {\bibinfo {booktitle} {Cold Molecules}}}\ (\bibinfo
  {publisher} {CRC Press},\ \bibinfo {year} {2009})\ pp.\ \bibinfo {pages}
  {453--502}\BibitemShut {NoStop}%
\bibitem [{\citenamefont {Peter}\ \emph {et~al.}(2012)\citenamefont {Peter},
  \citenamefont {M{\"u}ller}, \citenamefont {Wessel},\ and\ \citenamefont
  {B{\"u}chler}}]{peter2012anomalous}%
  \BibitemOpen
  \bibfield  {author} {\bibinfo {author} {\bibfnamefont {D.}~\bibnamefont
  {Peter}}, \bibinfo {author} {\bibfnamefont {S.}~\bibnamefont {M{\"u}ller}},
  \bibinfo {author} {\bibfnamefont {S.}~\bibnamefont {Wessel}}, \ and\ \bibinfo
  {author} {\bibfnamefont {H.~P.}\ \bibnamefont {B{\"u}chler}},\ }\href@noop {}
  {\bibfield  {journal} {\bibinfo  {journal} {Physical review letters}\
  }\textbf {\bibinfo {volume} {109}},\ \bibinfo {pages} {025303} (\bibinfo
  {year} {2012})}\BibitemShut {NoStop}%
\bibitem [{\citenamefont {Kadanoff}(2000)}]{kadanoff2000statistical}%
  \BibitemOpen
  \bibfield  {author} {\bibinfo {author} {\bibfnamefont {L.~P.}\ \bibnamefont
  {Kadanoff}},\ }\href@noop {} {\emph {\bibinfo {title} {Statistical physics:
  statics, dynamics and renormalization}}}\ (\bibinfo  {publisher} {World
  Scientific Publishing Company},\ \bibinfo {year} {2000})\BibitemShut
  {NoStop}%
\bibitem [{\citenamefont {Mellado}\ \emph {et~al.}(2012)\citenamefont
  {Mellado}, \citenamefont {Concha},\ and\ \citenamefont
  {Mahadevan}}]{mellado2012macroscopic}%
  \BibitemOpen
  \bibfield  {author} {\bibinfo {author} {\bibfnamefont {P.}~\bibnamefont
  {Mellado}}, \bibinfo {author} {\bibfnamefont {A.}~\bibnamefont {Concha}}, \
  and\ \bibinfo {author} {\bibfnamefont {L.}~\bibnamefont {Mahadevan}},\
  }\href@noop {} {\bibfield  {journal} {\bibinfo  {journal} {Physical review
  letters}\ }\textbf {\bibinfo {volume} {109}},\ \bibinfo {pages} {257203}
  (\bibinfo {year} {2012})}\BibitemShut {NoStop}%
\bibitem [{\citenamefont {Pollack}\ and\ \citenamefont
  {Stump}(2002)}]{pollack2002electromagnetism}%
  \BibitemOpen
  \bibfield  {author} {\bibinfo {author} {\bibfnamefont {G.~L.}\ \bibnamefont
  {Pollack}}\ and\ \bibinfo {author} {\bibfnamefont {D.~R.}\ \bibnamefont
  {Stump}},\ }\href@noop {} {\emph {\bibinfo {title} {Electromagnetism}}}\
  (\bibinfo  {publisher} {Addison-Wesley},\ \bibinfo {year} {2002})\BibitemShut
  {NoStop}%
\bibitem [{\citenamefont {McMichael}\ and\ \citenamefont
  {Donahue}(1997)}]{mcmichael1997head}%
  \BibitemOpen
  \bibfield  {author} {\bibinfo {author} {\bibfnamefont {R.~D.}\ \bibnamefont
  {McMichael}}\ and\ \bibinfo {author} {\bibfnamefont {M.~J.}\ \bibnamefont
  {Donahue}},\ }\href@noop {} {\bibfield  {journal} {\bibinfo  {journal} {IEEE
  Transactions on Magnetics}\ }\textbf {\bibinfo {volume} {33}},\ \bibinfo
  {pages} {4167} (\bibinfo {year} {1997})}\BibitemShut {NoStop}%
\bibitem [{\citenamefont {Tchernyshyov}\ and\ \citenamefont
  {Chern}(2005)}]{tchernyshyov2005fractional}%
  \BibitemOpen
  \bibfield  {author} {\bibinfo {author} {\bibfnamefont {O.}~\bibnamefont
  {Tchernyshyov}}\ and\ \bibinfo {author} {\bibfnamefont {G.-W.}\ \bibnamefont
  {Chern}},\ }\href@noop {} {\bibfield  {journal} {\bibinfo  {journal}
  {Physical review letters}\ }\textbf {\bibinfo {volume} {95}},\ \bibinfo
  {pages} {197204} (\bibinfo {year} {2005})}\BibitemShut {NoStop}%
\bibitem [{\citenamefont {Clarke}\ \emph {et~al.}(2008)\citenamefont {Clarke},
  \citenamefont {Tretiakov}, \citenamefont {Chern}, \citenamefont {Bazaliy},\
  and\ \citenamefont {Tchernyshyov}}]{clarke2008dynamics}%
  \BibitemOpen
  \bibfield  {author} {\bibinfo {author} {\bibfnamefont {D.}~\bibnamefont
  {Clarke}}, \bibinfo {author} {\bibfnamefont {O.}~\bibnamefont {Tretiakov}},
  \bibinfo {author} {\bibfnamefont {G.-W.}\ \bibnamefont {Chern}}, \bibinfo
  {author} {\bibfnamefont {Y.~B.}\ \bibnamefont {Bazaliy}}, \ and\ \bibinfo
  {author} {\bibfnamefont {O.}~\bibnamefont {Tchernyshyov}},\ }\href@noop {}
  {\bibfield  {journal} {\bibinfo  {journal} {Physical Review B}\ }\textbf
  {\bibinfo {volume} {78}},\ \bibinfo {pages} {134412} (\bibinfo {year}
  {2008})}\BibitemShut {NoStop}%
\bibitem [{\citenamefont {Tretiakov}\ \emph {et~al.}(2010)\citenamefont
  {Tretiakov}, \citenamefont {Liu},\ and\ \citenamefont
  {Abanov}}]{tretiakov2010minimization}%
  \BibitemOpen
  \bibfield  {author} {\bibinfo {author} {\bibfnamefont {O.~A.}\ \bibnamefont
  {Tretiakov}}, \bibinfo {author} {\bibfnamefont {Y.}~\bibnamefont {Liu}}, \
  and\ \bibinfo {author} {\bibfnamefont {A.}~\bibnamefont {Abanov}},\
  }\href@noop {} {\bibfield  {journal} {\bibinfo  {journal} {Physical review
  letters}\ }\textbf {\bibinfo {volume} {105}},\ \bibinfo {pages} {217203}
  (\bibinfo {year} {2010})}\BibitemShut {NoStop}%
\bibitem [{\citenamefont {Kapitza}(1951)}]{kapitza1951dynamic}%
  \BibitemOpen
  \bibfield  {author} {\bibinfo {author} {\bibfnamefont {P.}~\bibnamefont
  {Kapitza}},\ }\href@noop {} {\bibfield  {journal} {\bibinfo  {journal}
  {Soviet Physics--JETP}\ }\textbf {\bibinfo {volume} {21}},\ \bibinfo {pages}
  {588} (\bibinfo {year} {1951})}\BibitemShut {NoStop}%
\bibitem [{\citenamefont {Bar-Sinai}\ \emph {et~al.}(2020)\citenamefont
  {Bar-Sinai}, \citenamefont {Librandi}, \citenamefont {Bertoldi},\ and\
  \citenamefont {Moshe}}]{bar2020geometric}%
  \BibitemOpen
  \bibfield  {author} {\bibinfo {author} {\bibfnamefont {Y.}~\bibnamefont
  {Bar-Sinai}}, \bibinfo {author} {\bibfnamefont {G.}~\bibnamefont {Librandi}},
  \bibinfo {author} {\bibfnamefont {K.}~\bibnamefont {Bertoldi}}, \ and\
  \bibinfo {author} {\bibfnamefont {M.}~\bibnamefont {Moshe}},\ }\href@noop {}
  {\bibfield  {journal} {\bibinfo  {journal} {Proceedings of the National
  Academy of Sciences}\ }\textbf {\bibinfo {volume} {117}},\ \bibinfo {pages}
  {10195} (\bibinfo {year} {2020})}\BibitemShut {NoStop}%
\bibitem [{\citenamefont {Florijn}\ \emph {et~al.}(2014)\citenamefont
  {Florijn}, \citenamefont {Coulais},\ and\ \citenamefont {van
  Hecke}}]{florijn2014programmable}%
  \BibitemOpen
  \bibfield  {author} {\bibinfo {author} {\bibfnamefont {B.}~\bibnamefont
  {Florijn}}, \bibinfo {author} {\bibfnamefont {C.}~\bibnamefont {Coulais}}, \
  and\ \bibinfo {author} {\bibfnamefont {M.}~\bibnamefont {van Hecke}},\
  }\href@noop {} {\bibfield  {journal} {\bibinfo  {journal} {Physical review
  letters}\ }\textbf {\bibinfo {volume} {113}},\ \bibinfo {pages} {175503}
  (\bibinfo {year} {2014})}\BibitemShut {NoStop}%
\bibitem [{\citenamefont {Feynman}(1948)}]{RevModPhys.20.367}%
  \BibitemOpen
  \bibfield  {author} {\bibinfo {author} {\bibfnamefont {R.~P.}\ \bibnamefont
  {Feynman}},\ }\href {\doibase 10.1103/RevModPhys.20.367} {\bibfield
  {journal} {\bibinfo  {journal} {Rev. Mod. Phys.}\ }\textbf {\bibinfo {volume}
  {20}},\ \bibinfo {pages} {367} (\bibinfo {year} {1948})}\BibitemShut
  {NoStop}%
\bibitem [{\citenamefont {Kleinert}(2009)}]{kleinert2009path}%
  \BibitemOpen
  \bibfield  {author} {\bibinfo {author} {\bibfnamefont {H.}~\bibnamefont
  {Kleinert}},\ }\href@noop {} {\emph {\bibinfo {title} {Path integrals in
  quantum mechanics, statistics, polymer physics, and financial markets}}}\
  (\bibinfo  {publisher} {World scientific},\ \bibinfo {year}
  {2009})\BibitemShut {NoStop}%
\bibitem [{\citenamefont {Golubitsky}\ \emph {et~al.}(2012)\citenamefont
  {Golubitsky}, \citenamefont {Stewart},\ and\ \citenamefont
  {Schaeffer}}]{golubitsky2012singularities}%
  \BibitemOpen
  \bibfield  {author} {\bibinfo {author} {\bibfnamefont {M.}~\bibnamefont
  {Golubitsky}}, \bibinfo {author} {\bibfnamefont {I.}~\bibnamefont {Stewart}},
  \ and\ \bibinfo {author} {\bibfnamefont {D.~G.}\ \bibnamefont {Schaeffer}},\
  }\href@noop {} {\emph {\bibinfo {title} {Singularities and Groups in
  Bifurcation Theory: Volume II}}},\ Vol.~\bibinfo {volume} {69}\ (\bibinfo
  {publisher} {Springer Science \& Business Media},\ \bibinfo {year}
  {2012})\BibitemShut {NoStop}%
\bibitem [{\citenamefont {Doedel}\ \emph {et~al.}(1997)\citenamefont {Doedel},
  \citenamefont {Champneys}, \citenamefont {Fairgrieve}, \citenamefont
  {Kuznetsov}, \citenamefont {Sandstede}, \citenamefont {Wang} \emph
  {et~al.}}]{doedel1997continuation}%
  \BibitemOpen
  \bibfield  {author} {\bibinfo {author} {\bibfnamefont {E.~J.}\ \bibnamefont
  {Doedel}}, \bibinfo {author} {\bibfnamefont {A.~R.}\ \bibnamefont
  {Champneys}}, \bibinfo {author} {\bibfnamefont {T.~F.}\ \bibnamefont
  {Fairgrieve}}, \bibinfo {author} {\bibfnamefont {Y.~A.}\ \bibnamefont
  {Kuznetsov}}, \bibinfo {author} {\bibfnamefont {B.}~\bibnamefont
  {Sandstede}}, \bibinfo {author} {\bibfnamefont {X.}~\bibnamefont {Wang}},
  \emph {et~al.},\ }\href@noop {} {\bibfield  {journal} {\bibinfo  {journal}
  {AUTO97, Concordia University, Canada}\ } (\bibinfo {year}
  {1997})}\BibitemShut {NoStop}%
\bibitem [{\citenamefont {Parkin}\ \emph {et~al.}(2008)\citenamefont {Parkin},
  \citenamefont {Hayashi},\ and\ \citenamefont {Thomas}}]{parkin2008magnetic}%
  \BibitemOpen
  \bibfield  {author} {\bibinfo {author} {\bibfnamefont {S.~S.}\ \bibnamefont
  {Parkin}}, \bibinfo {author} {\bibfnamefont {M.}~\bibnamefont {Hayashi}}, \
  and\ \bibinfo {author} {\bibfnamefont {L.}~\bibnamefont {Thomas}},\
  }\href@noop {} {\bibfield  {journal} {\bibinfo  {journal} {Science}\ }\textbf
  {\bibinfo {volume} {320}},\ \bibinfo {pages} {190} (\bibinfo {year}
  {2008})}\BibitemShut {NoStop}%
\bibitem [{\citenamefont {Knobloch}(2015)}]{knobloch2015spatial}%
  \BibitemOpen
  \bibfield  {author} {\bibinfo {author} {\bibfnamefont {E.}~\bibnamefont
  {Knobloch}},\ }\href@noop {} {\bibfield  {journal} {\bibinfo  {journal}
  {Annu. Rev. Condens. Matter Phys}\ }\textbf {\bibinfo {volume} {6}},\
  \bibinfo {pages} {325} (\bibinfo {year} {2015})}\BibitemShut {NoStop}%
\bibitem [{\citenamefont {Plihon}\ \emph {et~al.}(2016)\citenamefont {Plihon},
  \citenamefont {Miralles}, \citenamefont {Bourgoin},\ and\ \citenamefont
  {Pinton}}]{plihon2016stochastic}%
  \BibitemOpen
  \bibfield  {author} {\bibinfo {author} {\bibfnamefont {N.}~\bibnamefont
  {Plihon}}, \bibinfo {author} {\bibfnamefont {S.}~\bibnamefont {Miralles}},
  \bibinfo {author} {\bibfnamefont {M.}~\bibnamefont {Bourgoin}}, \ and\
  \bibinfo {author} {\bibfnamefont {J.-F.}\ \bibnamefont {Pinton}},\
  }\href@noop {} {\bibfield  {journal} {\bibinfo  {journal} {Physical Review
  E}\ }\textbf {\bibinfo {volume} {94}},\ \bibinfo {pages} {012224} (\bibinfo
  {year} {2016})}\BibitemShut {NoStop}%
\bibitem [{\citenamefont {Nusse}\ and\ \citenamefont
  {Yorke}(1996)}]{nusse1996basins}%
  \BibitemOpen
  \bibfield  {author} {\bibinfo {author} {\bibfnamefont {H.~E.}\ \bibnamefont
  {Nusse}}\ and\ \bibinfo {author} {\bibfnamefont {J.~A.}\ \bibnamefont
  {Yorke}},\ }\href@noop {} {\bibfield  {journal} {\bibinfo  {journal}
  {Science}\ }\textbf {\bibinfo {volume} {271}},\ \bibinfo {pages} {1376}
  (\bibinfo {year} {1996})}\BibitemShut {NoStop}%
\bibitem [{\citenamefont {Grebogi}\ \emph {et~al.}(1986)\citenamefont
  {Grebogi}, \citenamefont {Kostelich}, \citenamefont {Ott},\ and\
  \citenamefont {Yorke}}]{grebogi1986multi}%
  \BibitemOpen
  \bibfield  {author} {\bibinfo {author} {\bibfnamefont {C.}~\bibnamefont
  {Grebogi}}, \bibinfo {author} {\bibfnamefont {E.}~\bibnamefont {Kostelich}},
  \bibinfo {author} {\bibfnamefont {E.}~\bibnamefont {Ott}}, \ and\ \bibinfo
  {author} {\bibfnamefont {J.~A.}\ \bibnamefont {Yorke}},\ }\href@noop {}
  {\bibfield  {journal} {\bibinfo  {journal} {Physics Letters A}\ }\textbf
  {\bibinfo {volume} {118}},\ \bibinfo {pages} {448} (\bibinfo {year}
  {1986})}\BibitemShut {NoStop}%
\bibitem [{\citenamefont {Menck}\ \emph {et~al.}(2013)\citenamefont {Menck},
  \citenamefont {Heitzig}, \citenamefont {Marwan},\ and\ \citenamefont
  {Kurths}}]{menck2013basin}%
  \BibitemOpen
  \bibfield  {author} {\bibinfo {author} {\bibfnamefont {P.~J.}\ \bibnamefont
  {Menck}}, \bibinfo {author} {\bibfnamefont {J.}~\bibnamefont {Heitzig}},
  \bibinfo {author} {\bibfnamefont {N.}~\bibnamefont {Marwan}}, \ and\ \bibinfo
  {author} {\bibfnamefont {J.}~\bibnamefont {Kurths}},\ }\href@noop {}
  {\bibfield  {journal} {\bibinfo  {journal} {Nature physics}\ }\textbf
  {\bibinfo {volume} {9}},\ \bibinfo {pages} {89} (\bibinfo {year}
  {2013})}\BibitemShut {NoStop}%
\bibitem [{\citenamefont {Grebogi}\ \emph {et~al.}(1983)\citenamefont
  {Grebogi}, \citenamefont {McDonald}, \citenamefont {Ott},\ and\ \citenamefont
  {Yorke}}]{grebogi1983final}%
  \BibitemOpen
  \bibfield  {author} {\bibinfo {author} {\bibfnamefont {C.}~\bibnamefont
  {Grebogi}}, \bibinfo {author} {\bibfnamefont {S.~W.}\ \bibnamefont
  {McDonald}}, \bibinfo {author} {\bibfnamefont {E.}~\bibnamefont {Ott}}, \
  and\ \bibinfo {author} {\bibfnamefont {J.~A.}\ \bibnamefont {Yorke}},\
  }\href@noop {} {\bibfield  {journal} {\bibinfo  {journal} {Physics Letters
  A}\ }\textbf {\bibinfo {volume} {99}},\ \bibinfo {pages} {415} (\bibinfo
  {year} {1983})}\BibitemShut {NoStop}%
\bibitem [{\citenamefont {Donahue}\ and\ \citenamefont
  {Donahue}(1999)}]{donahue1999oommf}%
  \BibitemOpen
  \bibfield  {author} {\bibinfo {author} {\bibfnamefont {M.~J.}\ \bibnamefont
  {Donahue}}\ and\ \bibinfo {author} {\bibfnamefont {M.}~\bibnamefont
  {Donahue}},\ }\href@noop {} {\emph {\bibinfo {title} {OOMMF user's guide,
  version 1.0}}}\ (\bibinfo  {publisher} {US Department of Commerce, National
  Institute of Standards and Technology},\ \bibinfo {year} {1999})\BibitemShut
  {NoStop}%
\bibitem [{\citenamefont {Daza}\ \emph {et~al.}(2016)\citenamefont {Daza},
  \citenamefont {Wagemakers}, \citenamefont {Georgeot}, \citenamefont
  {Gu{\'e}ry-Odelin},\ and\ \citenamefont {Sanju{\'a}n}}]{daza2016basin}%
  \BibitemOpen
  \bibfield  {author} {\bibinfo {author} {\bibfnamefont {A.}~\bibnamefont
  {Daza}}, \bibinfo {author} {\bibfnamefont {A.}~\bibnamefont {Wagemakers}},
  \bibinfo {author} {\bibfnamefont {B.}~\bibnamefont {Georgeot}}, \bibinfo
  {author} {\bibfnamefont {D.}~\bibnamefont {Gu{\'e}ry-Odelin}}, \ and\
  \bibinfo {author} {\bibfnamefont {M.~A.}\ \bibnamefont {Sanju{\'a}n}},\
  }\href@noop {} {\bibfield  {journal} {\bibinfo  {journal} {Scientific
  reports}\ }\textbf {\bibinfo {volume} {6}},\ \bibinfo {pages} {1} (\bibinfo
  {year} {2016})}\BibitemShut {NoStop}%
\bibitem [{\citenamefont {Cambr{\'e}}\ \emph {et~al.}(2015)\citenamefont
  {Cambr{\'e}}, \citenamefont {Campo}, \citenamefont {Beirnaert}, \citenamefont
  {Verlackt}, \citenamefont {Cool},\ and\ \citenamefont
  {Wenseleers}}]{cambre2015asymmetric}%
  \BibitemOpen
  \bibfield  {author} {\bibinfo {author} {\bibfnamefont {S.}~\bibnamefont
  {Cambr{\'e}}}, \bibinfo {author} {\bibfnamefont {J.}~\bibnamefont {Campo}},
  \bibinfo {author} {\bibfnamefont {C.}~\bibnamefont {Beirnaert}}, \bibinfo
  {author} {\bibfnamefont {C.}~\bibnamefont {Verlackt}}, \bibinfo {author}
  {\bibfnamefont {P.}~\bibnamefont {Cool}}, \ and\ \bibinfo {author}
  {\bibfnamefont {W.}~\bibnamefont {Wenseleers}},\ }\href@noop {} {\bibfield
  {journal} {\bibinfo  {journal} {Nature nanotechnology}\ }\textbf {\bibinfo
  {volume} {10}},\ \bibinfo {pages} {248} (\bibinfo {year} {2015})}\BibitemShut
  {NoStop}%
\bibitem [{\citenamefont {Mellado}\ \emph {et~al.}(2020)\citenamefont
  {Mellado}, \citenamefont {Concha},\ and\ \citenamefont
  {Rica}}]{PhysRevLett.125.237602}%
  \BibitemOpen
  \bibfield  {author} {\bibinfo {author} {\bibfnamefont {P.}~\bibnamefont
  {Mellado}}, \bibinfo {author} {\bibfnamefont {A.}~\bibnamefont {Concha}}, \
  and\ \bibinfo {author} {\bibfnamefont {S.}~\bibnamefont {Rica}},\ }\href
  {\doibase 10.1103/PhysRevLett.125.237602} {\bibfield  {journal} {\bibinfo
  {journal} {Phys. Rev. Lett.}\ }\textbf {\bibinfo {volume} {125}},\ \bibinfo
  {pages} {237602} (\bibinfo {year} {2020})}\BibitemShut {NoStop}%
\end{thebibliography}
\end{document}